\documentclass[pre,showpacs,preprintnumbers ]{revtex4}
\usepackage{graphicx}
\usepackage{epstopdf}
\usepackage{amsmath}
\usepackage{amssymb}
\usepackage{mathrsfs}
\usepackage[mathscr]{eucal}
\usepackage{bm}% bold math

\def\ie{i.e.}
\def\eg{e.g.}

\def\half{\frac{1}{2}}
\def\smallhalf{\tfrac{1}{2}}

\def\roottwo \sqrt{2}

\providecommand{\abs}[1]{\lvert#1\rvert}
\providecommand{\unitvec}[1]{\hat{\boldsymbol{#1}}} % unit vector format
\providecommand{\cc}{^{\ast}}

\providecommand{\del}{\partial}

\def\be{\begin{equation}}
\def\ee{\end{equation}}
\def\beu{\begin{equation*}}
\def\eeu{\end{equation*}}

\begin{document}

% title page

\preprint{UCB/LBNL-CBP preprint}

\title{Robust autoresonant excitation in the plasma beat-wave accelerator:\\ a theoretical study}
\author{R.R. Lindberg}
\email{RL236@socrates.berkeley.edu}
\author{A.E. Charman}
\author{J.S. Wurtele}
\affiliation{Department of Physics, University of California, Berkeley, Berkeley, CA 94720, USA\\
Center for Beam Physics, Lawrence Berkeley National Laboratory, USA}
\author{L. Friedland}
\affiliation{Racah Institute of Physics, The Hebrew University, Jerusalem 91904, Israel}
\date{15 March 2004}

% abstract %
\begin{abstract}
A modified version of the Plasma Beat-Wave Accelerator scheme is introduced and analyzed, which  is based on autoresonant phase-locking of the nonlinear Langmuir wave to the slowly chirped beat frequency of the driving lasers via adiabatic passage through resonance.  This new scheme  is designed to overcome some of the well-known limitations of previous approaches, namely relativistic detuning and nonlinear modulation or other non-uniformity or non-stationarity in the driven Langmuir wave amplitude, and sensitivity to frequency mismatch due to measurement uncertainties and density fluctuations and inhomogeneities.  As in previous schemes, modulational instabilities of the ionic background ultimately limit the useful interaction time, but nevertheless peak electric fields at or approaching the wave-breaking limit seem readily attainable.  Compared to traditional approaches, the autoresonant scheme achieves larger accelerating electric fields for given laser intensity, or comparable fields for less laser power;  the plasma wave excitation is much more robust to  variations or uncertainties in plasma density; it is largely insensitive to the precise choice of chirp rate, provided only that chirping is sufficiently slow; and the quality and uniformity of the resulting  plasma wave and its suitability for accelerator applications may be superior.  In underdense plasmas, the total frequency shift required is only of the order of a few percent of the laser carrier frequency, and for possible experimental proofs-of-principle, the scheme might be implemented with relatively little additional modification to existing systems based on either solid-state amplifiers and Chirped Pulse Amplification techniques, or, with somewhat greater technological effort, using a $\mbox{CO}_{2}$ or other gas laser system.
\end{abstract}
\maketitle

% introduction .....

\setlength{\parindent}{0em}
\setlength{\parskip}{1.75ex plus 0.5ex minus 0.2ex}

\section{Introduction and Overview}\label{section:introduction}

The Plasma Beat-Wave Accelerator (PBWA) was first proposed by Tajima and Dawson (TD) \cite{tajima_dawson:79} as an alternative to the short-pulse Laser Wake-Field Accelerator (LWFA), based on earlier analysis of  beat-wave excitation as a plasma probe \cite{kroll_et_al:64} or as a mechanism for plasma heating \cite{kroll_et_al:64, cohen_et_al:72, rosenbluth_liu:72}.  Subsequently the PBWA concept has been studied extensively theoretically, numerically, and experimentally \cite{joshi_et_al:84, amini_chen:84, tang_et_al:84, tang_et_al:85, forslund_kindel:85, clayton_et_al:85, noble:85, mori:87, mora_et_al:88,  dangor_et_al:90, kitigawa_et_al:92, clayton_et_al:93, everett_et_al:94, clayton_et_al:98}.  (For a review, see \cite{esarey_et_al:96}.)  In the original scheme, two laser fields co-propagating in an underdense plasma are detuned from each other by a frequency shift close to the electron plasma frequency, so that the modulated envelope resulting from the beating between the two laser fields can act ponderomotively on the plasma electrons to resonantly excite a large-amplitude, high-phase-velocity plasma wave suitable for particle acceleration.

For a fixed beat frequency, performance of the PBWA is constrained by what is now known as the Rosenbluth-Liu (RL) limit, after the pioneering study in \cite{rosenbluth_liu:72}.  As the plasma wave amplitude grows, relativistic detuning effects eventually prevent further growth of the peak longitudinal electric field $E_{z}$ beyond a maximum value $E_{\text{\tiny{RL}}},$ which, for any realistic laser and plasma parameters, lies below the cold, non-relativistic, one-dimensional (1D), wave-breaking limit $E_{0}$ \cite{akhiezer_polovin:56, dawson:59}:
\be
\abs{E_{z}} \le E_{\text{\tiny{RL}}} \equiv   E_{0} \left(\tfrac{16}{3}\tfrac{\omega_p^2}{\omega_1\omega_2} \tfrac{\abs{E_1}\abs{E_2}}{E_0^2} \right)^{1/3} < E_{0} \equiv \frac{m c \omega_{p}}{e},
\ee
where $c$ is the speed of light, $m$ is the mass of the electron, $e$ is the magnitude of its electric charge; 
$\omega_p$ is the cold, linear electron plasma frequency, defined in Gaussian units as
\be
\omega_{p} = \left(4\pi n_0 e^2/m \right)^{1/2},
\ee
where $n_0$ is the ambient electron number density; $E_1$ and $E_2$ are the peak electric fields of the beating drive lasers; and $\omega_1$ and $\omega_2$ are their respective carrier frequencies.
For the plasma waves of interest here, with high but sub-luminal phase velocities $v_{p} \lesssim c,$ the cold non-relativistic limit itself is smaller than the cold, relativistic (or nonlinear) wave-breaking limit \cite{akhiezer_polovin:56, esarey_et_al:96}:
 \be\label{eqn:wavebreaking}
 E_0 < E_{\text{\tiny{WB}}} = \left[2(\gamma_{p} - 1) \right]^{1/2}E_{0} ,
 \ee
 which sets an upper bound on the amplitude of coherent plasma oscillations useful for particle acceleration in a cold plasma.  Here
 \be
 \gamma_p = \left[1 - v_p^2/c^2 \right]^{-1/2}
 \ee
 is the relativistic kinematic factor associated with the phase-velocity $v_p$ of the excited plasma wave, approximately equal to the characteristic group velocity $\bar{v}_g$ (to be precisely defined below) associated with the envelope of the beating lasers, both of which will be nearly equal to the speed of light $c$ for sufficiently underdense plasmas.  (Large thermal spreads increase particle trapping and lower this threshold for wave-breaking, as discussed in \cite{katsouleas_mori:88, rosenzweig:88, mori_katsouleas:90}.)
   
It is not difficult to see the origin of this detuning effect.  The effective nonlinear plasma frequency, in the presence of the laser drive and a plasma wave, is approximately given by 
 \be
 \omega_{p_\text{\tiny{NL}}} \approx  \frac{\left(1 + \overline{\delta n}/n_0 \right)}{\sqrt{\gamma_{\text{\tiny{RMS}}}}}\omega_{p},
 \ee
 where $\overline{\delta n}$ is an average density perturbation associated with the plasma wave, and $\gamma_{\text{\tiny{RMS}}}$ is the root-mean-square relativistic factor of the electrons, including effects of both the transverse quiver in laser fields and the longitudinal velocity imparted by the Langmuir wave itself.   Except for highly nonlinear situations where ponderomotive blow-out becomes important,  the average density perturbation is typically small, and will be neglected here.  The RMS transverse quiver velocity is roughly constant for the fixed-amplitude driving lasers typical of PBWA, but as the longitudinal motion of the perturbed electrons associated with the excited Langmuir wave becomes even weakly relativistic, $\gamma_{\text{\tiny{RMS}}}$ will begin to increase, causing  $\omega_{p_\text{\tiny{NL}}}$ to decrease.  Because of the sensitivity of resonance phenomena, $\gamma_{\text{\tiny{RMS}}}$ need not grow much before this resonance shift severely limits the efficacy of the driving beat-wave.  As this dynamical process is effectively reversible, if the detuning occurs before wave-breaking or other phase-mixing or dissipative effects, then the ponderomotively excited Langmuir wave is not actually saturated \textit{per se} at some fixed value given by the RL limit, but exhibits a slow (compared to the plasma frequency) nonlinear modulation or beating of the wave amplitude, periodically peaking near the maximum value predicted by RL and then decreasing, as energy is exchanged between the Langmuir wave and the laser fields.
 
Tang, Sprangle, and Sudan (TSS) \cite{tang_et_al:84, tang_et_al:85} pointed out that one can, in principle, achieve somewhat higher amplitudes than that predicted by the RL limit by simply detuning the beat frequency to a value somewhere below the linear plasma frequency, matching the drive frequency to the nonlinear plasma frequency for some non-trivial plasma wave amplitude.  However, McKinstrie and Forslund \cite{mckinstrie_forslund:87} noted fundamental practical difficulties with this approach.  For downward detunings at or beyond some small critical frequency shift, multiple solutions to the envelope equations appear, and the high-amplitude branch cannot be accessed reliably from the initially quiescent plasma state, especially in the presence of any small amount of damping.  For sufficiently small detunings, moderate improvement over the RL limit is possible, but typically with peak amplitudes still lying below even the cold linear wave-breaking limit.   Furthermore, as the detuning grows in magnitude toward this critical value, the excitation becomes increasingly non-monotonic, and the time needed to reach the peak amplitude increases and becomes comparable to the time-scales for ion modulational or decay instabilities that degrade the coherence of the plasma wave together with its utility for accelerator applications.

Matte \textit{et al.} (MMEBP) \cite{matte_et_al:87} suggested the use of a plasma with a time-varying density, carrying out the beat-wave excitation during the actual ionization process, when the free electron density is growing.  The increase in plasma density  can then compensate for the increase in $\gamma_{\text{\tiny{RMS}}},$ and thus the effective plasma resonance may be maintained with a fixed driving beat frequency for a longer time.  However, in order to yield appreciable improvement, such an approach would require impractically precise timing of lasers and some means to control the ionization rate, matching it, at least roughly, to the expected growth rate of the plasma wave.  The usual method involving laser-induced field ionization leads to cross sections which are exponentially sensitive to the laser intensity and is therefore unsuitable for such an approach.

Better still, Deutsch, Meerson, and Golub (DMG) \cite{deutsch_et_al:91} suggested incorporating compensatory time-dependence into the drive lasers rather than in the plasma density.  They proposed to partially overcome the relativistic detuning effect by using a chirped beat-wave excitation scheme, where the frequencies of one or more of the laser pulses are chirped downward starting from the linear resonance so as to compensate for the change in the nonlinear  frequency of the growing plasma wave, ideally allowing amplitudes approaching the nonlinear wave-breaking limit.  The nonlinear modulations are thereby reduced but not eliminated, consisting of a ringing about some non-zero saturation value rather than a full beating as in the original (RL/TD) scheme.

Chirping too fast will cause the plasma oscillations to detune before they have a chance to grow much, while chirping too slowly appears to result in a slow, highly non-monotonic (ringing) excitation, where the long interaction times required to build up sufficient amplitude (on the order of a few ion plasma periods) can allow instabilities of the ionic background to grow and disrupt the coherence of the plasma wave \cite{darrow_et_al:87, mora_et_al:88}.  Intuitively, one expects for the DMG approach that there exists an optimal intermediate chirp rate, or in fact even an entire optimal chirp profile, in which the driving beat frequency starts at the linear plasma resonance, where the initial coupling is strongest, and then self-consistently matches the instantaneous frequency to the detuning caused by the expected growth of the plasma wave.

However,  solving such a complicated nonlinear control problem so as to exactly match the nonlinear plasma frequency with the laser chirp, with the help of either numerical simulation or experimental calibration, would be both computationally daunting and experimentally impractical.  The linear plasma frequency will be imperfectly known in practice, and even if it were obtained, the instantaneous value of the nonlinear plasma frequency actually depends on the initial absolute phase of the plasma wave, which is determined by the exact initial conditions of the plasma and the initial beat phase of the driving lasers, which are not normally controlled.  However, DMG argue that such {\it extrinsic} frequency matching is not actually necessary, and that entrainment can be achieved without recourse to carefully-tailored pulses or to external feedback.  They invoke the nonlinear dynamical phase-locking phenomenon known as autoresonance \cite{loeb_friedland:86, meerson_friedland:90}, arguing that as long as the chirp is sufficiently slow (compared to the time-scales for the nonlinear growth and modulation described by RL) and in the right direction (namely, downward in frequency), then the externally forced, nonlinear dynamical system can self-adjust and automatically maintain approximately the desired phase entrainment between the driven wave and the ponderomotive drive, resulting in in an average increase in the oscillation amplitude.

However, proposed as it was before the understanding of autoresonance had matured, especially with regard to the threshold behavior for establishing and maintaining entrainment and the importance of the initial detuning, the DMG scheme sometimes fails to produce appreciable phase-locking.  This is largely due to its prescription of beginning the chirp on resonance, and becomes particularly problematic if the plasma density is imperfectly known or subject to shot-to-shot jitter, when it can lead to nonuniform growth or early saturation.  In fact, there appears to be some uncertainty in the original description of the scheme \cite{deutsch_et_al:91} as to exactly how autoresonant the proposed excitation mechanism is, as to whether or to what extent ``the precise form of the time-dependence of the frequency is inessential'' or else there ``exists an optimal chirp rate that provides the highest excitation rate.''

In fact  the DMG chirped scheme can often achieve higher longitudinal electric fields than the original (RL/TD) approach, but does not do so universally or robustly, and cannot always produce fields approaching even the linear wave-breaking limit for experimentally accessible parameter values.  The peak amplitude and spatio-temporal uniformity of the final plasma wave may be sensitive to the exact chirp rate, to the initial detuning, and most especially to the precise value of the plasma frequency, which, once again, often is imperfectly known or subject to significant variation, due to limited measurement precision, to shot-to-shot jitter, or to single-shot density fluctuations.

Informed by more recent results, we propose a novel variant of the chirped plasma PBWA concept which also exploits autoresonance, but enhanced via Adiabatic Passage Through Resonance (APTR) \cite{friedland:92, friedland:97, friedland:98}.  Rather than chirping downward from the linear resonance, we actually start, counter to naive intuition, with a frequency shift well above the resonance, and then slowly sweep the beat frequency through and below resonance.  With this approach, when starting from a quiescent plasma, the final state of the plasma wave is insensitive to the exact chirp history, and the excitation is more robust with respect to imprecise characterization of the plasma density, or actual shot-to-shot fluctuation of the density or single-shot variation in it, provided only that the spatio-temporal scales of variability are long compared to the frequency and wavenumber of the plasma wave, and the relative range of fluctuation or uncertainty is not too large.  The chirp rate need not be matched \textit{a priori} to any anticipated rate of growth in the plasma wave or resulting rate of relativistic detuning, but must only be chosen to be sufficiently slow so as to satisfy a certain adiabatic trapping condition and thereby ensure phase-locking.  Because of the nonlinear nature of the ponderomotive forcing in the relativistic regime, this adiabaticity condition becomes increasingly difficult to maintain as the wave amplitude grows, and for a constant chirp rate eventually the plasma wave would fall out of phase-locking with the drive.  This leads to a true saturation in amplitude, with significantly less of the undesirable ringing seen in the original RL/TD scheme.  Before the onset of saturation, the frequency of the excited plasma wave can be closely entrained to the instantaneous beat frequency of the drivers, and the overall phase can remain reasonably entrained as well, while the plasma wave amplitude monotonically grows  to automatically and self-consistently adjust itself to the monotonically decreasing beat frequency.

Not only is our excitation scheme more robust, so too is our method of analysis.  The Lagrangian fluid formalism developed by RL and then used by DMG employs a power series expansion which is valid only for weakly relativistic motion in both the transverse and longitudinal directions, thereby limiting its domain of applicability to laser fields of sufficiently low intensity (implying non-relativistic quiver motion), and to excited plasma waves of sufficiently low amplitude (with electric fields well below wave-breaking).  Therefore their equations of motion become increasingly untrustworthy precisely in the regime of interest, where the plasma wave amplitude grows to some appreciable fraction of the wave-breaking value $E_0.$  Here, we instead employ a fully nonlinear, Eulerian fluid model that allows for arbitrarily relativistic electron motion below the nonlinear wave-breaking limit $E_{\text{\tiny{WB}}},$ similar to that discussed in \cite{sprangle_et_al:90, esarey_et_al:96} for general laser-plasma interactions and in \cite{noble:85} for PBWA investigations, and also to a pioneering treatment  in \cite{akhiezer_polovin:56}.  In the weakly relativistic limit, this Eulerian approach can be shown to agree exactly with the Lagrangian treatment of RL.

This analytical fluid model, and the various physical assumptions which go into it, are discussed in Section \ref{section:equations}.  In order to treat the autoresonant nature of the problem more easily and transparently, we re-formulate the dynamical equations in a fully Hamiltonian form in Section \ref{section:hamiltonian}.  In Section \ref{section:autoresonance}, we use this canonical formalism to analyze the autoresonant aspects of the beat-wave excitation, from the initial linear phase-locking regime through the weakly and strongly nonlinear trapped phases to non-adiabatic saturation.  In Section \ref{section:experimental}, we discuss features of some realistic examples relevant for possible experimental implementation and investigation.  Section \ref{section:discussion} summarizes our preliminary assessment of the merits and limits of our autoresonant PBWA, especially in comparison to other beat-wave approaches.   We then offer brief conclusions from our initial investigation and prospects for future study in Section \ref{section:conclusion}.
 
%    section on: mathematical model

\section{Fundamental Equations}\label{section:equations}

Our study of wake excitation is based on an approximate, but convenient and widely-used model.
The plasma is treated as a cold, collisionless, fully relativistic electron fluid moving in a stationary, neutralizing, ionic background, coupled to electromagnetic fields governed by Maxwell's equations.  We restrict our analysis to one-dimensional (1D) geometry, where all dynamical quantities depend only on the longitudinal position $z$ and the time $t,$ and we assume a completely  homogeneous  and initially quiescent background state of the plasma.  The plasma is assumed to be highly underdense, \ie, $\omega_p \ll \omega_1,\, \omega_2.$  In addition, we assume prescribed laser fields, neglecting throughout the entire interaction any changes to the laser envelopes due to linear effects such as diffraction or group-velocity dispersion, or any nonlinear back-action of the plasma on the lasers such as depletion, self-focusing, photon acceleration due to ionization or density variation, as well as Raman scattering, self-modulation, and other instabilities \cite{forslund_et_al:73, max_et_al:74}.  This model, although simplified, nevertheless reveals the essential features of autoresonance and its potential advantages for the PBWA.  Possible extensions to more realistic models, as well as some arguments for the validity of our general results beyond the strict applicability of this model, especially in the light of density fluctuations, are discussed in Sec.~\ref{section:discussion} and Sec.~\ref{section:conclusion}.

The cold, collisionless fluid treatment assumes that the electron temperature is sufficiently small so that:  the thermal energy $k_{\text{B}}T_{e}$ is negligible compared to the typical kinetic energy associated with the transverse quiver motion in the driving laser fields; thermal corrections to the Langmuir dispersion relation are small, which, for waves with relativistic phase velocities simply requires that $k_{\text{B}}T_{e}$ is much smaller than the electron rest energy $mc^2;$ and collisional damping of the laser fields is small over the interaction time, as are Landau and collisional damping of the excited plasma wave.  Neglect of ion dynamics is strictly valid provided the time-scale for ion motion, typically of the order of a few times $\omega_{i}^{-1},$ where 
\be
\omega_{i} = \left(4\pi n_0 Z_{i}^2e^2/M_{i} \right)^{1/2}
\ee
 is the ion plasma frequency for ions of mass $M_{i}$ and charge $+Z_{i}e,$
remains longer than the duration $T$ of the lasers:  \ie, $\omega_{i}T \lesssim 2\pi.$

Within this model, the continuity, momentum, and Poisson equations, respectively, are, in Gaussian units:
\begin{eqnarray}
  \frac{\partial n_e}{\partial t} + \boldsymbol{\nabla}\cdot
	(n_e\boldsymbol{v}) &=& \phantom{-}0 \label{eqn:continuity}\\
  \frac{\partial \boldsymbol{p}}{\partial t} + (\boldsymbol{v}\cdot
	\boldsymbol{\nabla})\boldsymbol{p} &=& \phantom{-}e\left(\boldsymbol{\nabla}\Phi - 
	\frac{\boldsymbol{v}}{c}\times\left(\boldsymbol{\nabla}\times\boldsymbol{A}\right)\right)  \label{eqn:momentum}\\
  \nabla^2\Phi &=&  -4\pi e(n_0-n_e) \label{eqn:poisson},
\end{eqnarray}
where $\boldsymbol{A}$ is the vector potential and $\Phi$ the scalar potential in the Coulomb gauge;  $n_e$ is the electron number density and $n_0$ the background ion number density, assumed to be equal in the absence of perturbations; $\boldsymbol{v}$ is the electron velocity; and $\boldsymbol{p} = m \gamma \boldsymbol{v}$ is the kinetic momentum, where
\be
\gamma = \left[1 - \abs{\boldsymbol{v}}^2/c^2 \right]^{-1/2}
\ee
is the relativistic factor associated with electron motion.

Neglect of diffraction or other transverse effects in the laser fields requires that characteristic laser spot size $w_0$ sufficiently exceeds both the characteristic laser wavelength $\lambda$ and the transverse length-scale $\Delta r_{\perp}$ for variations in the plasma, while $\Delta r_{\perp}$  itself should greatly exceed the Langmuir wavelength $\lambda_p \sim c/\omega_p$ to ensure the validity of our assumption of a 1D, homogeneous medium.  Neglect of the nonlinear evolution of the lasers effectively imposes certain constraints on the duration and intensity of the pulses, as summarized in \cite{esarey_et_al:96}.   In the Coulomb gauge, the vector potential $\boldsymbol{A}$ is solenoidal (\ie, divergenceless), which in 1D also implies that it is geometrically transverse, so $\boldsymbol{A} = \boldsymbol{A}_{\perp}(z, t)$ will be polarized perpendicular to the propagation direction $+\unitvec{z}.$   For simplicity each drive laser will be assumed to possess right-handed circular polarization, and to have a flat-topped profile of total duration $T,$ where the leading edge of each laser enters the left edge of the plasma at $z = 0$ at the initial time $t = 0.$

Defining a scaled, or dimensionless, time $\tau \equiv \omega_p t,$ co-moving position $\xi \equiv 
\omega_p(t - z/\bar{v}_g)$,  longitudinal velocity $\beta \equiv v_{z}/c,$ Langmuir phase velocity $\beta_p \equiv v_{p}/c = \bar{v}_g/c,$  vector potential $\boldsymbol{a} = \boldsymbol{a}_{\perp}  \equiv 
\frac{e}{mc^2}\boldsymbol{A}$, electrostatic potential $\phi \equiv 
\frac{e}{mc^2}\Phi$, and number density perturbation $\rho \equiv \left(n_e - n_0\right)/n_0$, the continuity equation (\ref{eqn:continuity}) can be written in scaled variables as
\be\label{eqn:continuity_scaled}
\frac{\del}{\del \tau} \rho + \frac{\del}{\del \xi}\left[\left(1 -
\beta/\beta_{p}\right)\left(1 + \rho \right) \right] = 0,
\ee
and Poisson's equation (\ref{eqn:poisson}) becomes
\be\label{eqn:poisson_scaled}
\frac{\del^{2}}{\del \xi^{2}}\phi = \beta_{p}^{2} \rho.
\ee
Assuming an initially quiescent plasma in the initial absence of laser fields (\ie, $\eta,\, \phi, \, \beta,\, \boldsymbol{a} \to  0$ as $\tau \to -\infty$ for fixed $\xi$ or as $\xi \to -\infty$ for fixed $\tau$), the conservation of transverse canonical momentum implies that
\be
\boldsymbol{p}_\perp = mc\boldsymbol{a}_\perp = mc\boldsymbol{a}. 
\ee
The relativistic factor $\gamma$ then factors into transverse and longitudinal contributions as
\be\label{eqn:gamma_factored}
\gamma \equiv \gamma_{\perp}\gamma_{\parallel} = \sqrt{1+ \abs{\boldsymbol{a}}^2}\frac{1}{\sqrt{1 -\beta^{2}}}.
\ee
Using these relations, we eliminate any explicit appearance of $\boldsymbol{v}_{\perp}$ or $\boldsymbol{p}_{\perp}$ in the longitudinal component of the momentum equation (\ref{eqn:momentum}), thereby obtaining
\be\label{eqn:momentum_scaled}
\frac{\del}{\del \tau}(\gamma\beta) + \left(1 -\beta/\beta_{p}\right)\frac{\del}{\del \xi}(\gamma\beta) =
-\frac{1}{\beta_{p}}\frac{\del}{\del \xi}\phi + \frac{1}{2}\frac{1}{\beta_{p}\gamma}\frac{\del}{\del \xi}\abs{\boldsymbol{a}}^{2}.
\ee

Now, we further make the Quasi-Static Approximation (QSA) (see, \eg, \cite{noble:85} and 
\cite{sprangle_et_al:91}), wherein we assume $\rho = \rho(\xi)$ and $\beta = 
\beta(\xi)$, \ie,  that the plasma response is independent of time $\tau$ in the 
co-moving frame, implying that the plasma wave itself moves without dispersion at the fixed
average group velocity $\beta_p$ of the driving lasers.  The QSA also specifically requires that any distortion or change in the shape of the laser envelope remains negligible during the typical interaction time with any one transverse slice of plasma, which will be $O(T).$   This constraint is satisfied automatically in our model,  since we have already assumed that laser envelope evolution is negligible throughout the entire interaction, which is of duration $O(T +  L/c),$ where $L$ is the total plasma length.  In the QSA, all $\tau$ derivatives in equations describing the plasma response can then be neglected, and the continuity equation (\ref{eqn:continuity_scaled}) can be immediately integrated, yielding, for our quiescent initial conditions,
\be\label{eqn:qsa_integral1}
\left(1 - \beta/\beta_{p}\right)\left( 1 + \rho \right) = 1.
\ee
Expressing $\abs{\boldsymbol{a}}^{2}$ in terms of $\gamma$ and $\beta,$ the longitudinal momentum equation (\ref{eqn:momentum_scaled}) can also be integrated to obtain:
\be\label{eqn:qsa_integral2}
\gamma\left(1 - \beta_{p}\beta\right) - \phi = 1.
\ee
After some algebra, the Poisson equation (\ref{eqn:poisson_scaled}) then becomes
\begin{equation}\label{eqn:quasi_mess}
\frac{\partial^{2}}{\partial \xi^{2}}\phi = \beta_p^2\gamma_{p}^{2}\left[\beta_{p}\left(1 -
\frac{1+ \abs{\boldsymbol{a}}^{2}}{\gamma_{p}^{2}(1+\phi)^{2}}\right)^{-\frac{1}{2}} - 1\right].
\end{equation}
This single, second-order equation describing the nonlinear plasma response is valid, within the QSA, for arbitrary relativistic electron velocities provided the Langmuir wave amplitude remains below the nonlinear wave-breaking limit determined by:
\be\label{eqn:qsa_wb}
\phi > -1 + \frac{1}{\gamma_{p}}\sqrt{1 + \abs{\boldsymbol{a}}^{2}} = -1 + \frac{\gamma_{\perp}}{\gamma_{p}}.
\ee
In the high phase-velocity limit appropriate to highly underdense plasmas, where $0 < 1- \beta_p \ll 1$ and $\gamma_{p} \gg 1$, the dynamical equation (\ref{eqn:quasi_mess}) may be Taylor expanded and further simplified to
\begin{equation}\label{eqn:quasi_simpler}
 \frac{\partial^2}{\partial\xi^2} \phi = \frac{1}{2}\left[\frac{1+a^{2}}{(1+\phi)^{2}} -1\right].
\end{equation}
Because we have effectively taken $\gamma_p \to \infty,$ this equation remains mathematically well-defined for any $\phi > -1$ and any $\abs{\boldsymbol{a}}^{2} \ge 0,$ but the actual bound  (\ref{eqn:qsa_wb}) remains a more physically trustworthy limit for wave-breaking. After solving for $\phi(\xi),$ we can determine the electric field using
\be
\frac{\partial}{\partial\xi} \phi = \beta_p \varepsilon_{\|}
\ee
where $\varepsilon_{\|} = \varepsilon_{\|}(\xi) \equiv E_z/E_0$ is the scaled longitudinal electric field within the QSA.  
The density perturbation $\rho(\xi)$ may then be determined using the scaled Poisson's equation (\ref{eqn:poisson_scaled}), the velocity $\beta(\xi)$ using the first integral (\ref{eqn:qsa_integral1}), and finally $\gamma(\xi)$ by using (\ref{eqn:gamma_factored}).

Under our assumptions, the normalized vector potential 
\be
\boldsymbol{a} = \boldsymbol{a}(\xi, \tau)  = \frac{e}{mc^{2}}\boldsymbol{A}\left(z = (\tau - \xi)\bar{v}_g/\omega_p, t = \tau/\omega_p \right)
\ee
is taken to be a prescribed function of time and position, describing the unperturbed laser fields of duration $T$ throughout the plasma of length $L,$  with a modulated envelope which is slowly chirped but travels along at a fixed group velocity $\bar{v}_g.$  For otherwise arbitrary laser fields, $\abs{\boldsymbol{a}}^{2}$  will still have both fast (carrier frequency) and slow (beat frequency) dependence on both $\tau$ and $\xi,$ although considerable simplification is possible for the assumed PBWA form, namely: two weakly detuned, slowly-chirped, co-propagating, nearly-plane-wave, flat-topped lasers with positive helicity.  The normalized vector potential $\boldsymbol{a},$ representing the beating drive lasers may then be explicitly written as
\be\label{eqn:laser_fields}
\boldsymbol{a} = \boldsymbol{a}_1 + \boldsymbol{a}_2  = \smallhalf \left[ \unitvec{e}_{+}a_1 e^{i\psi_1} + c.c.\right] + 
\smallhalf \left[ \unitvec{e}_{+}a_2 e^{i\psi_2} +  c.c.\right],
\ee
where $\unitvec{e}_{+} \!= \tfrac{1}{\sqrt{2}}\left(\unitvec{x} + i \unitvec{y} \right).$

At the leading edge ($z = 0$) of the plasma, the laser phases $\psi_{j}$ ($j = 1, 2$) are given, for $t \ge 0,$ by
\be
\psi_j(z = 0, t) = \psi_{j_0} -\int\limits_{0}^{t} d t' \, \omega_j(t'),
\ee
where the $\psi_{j_0}$ are real constants depending on initial laser conditions, and the instantaneous carrier frequencies $\omega_j(t)  \equiv \omega_j( z = 0, t) \equiv  -\frac{d}{d t}\psi_{j}( z = 0, t)$ allow for slow chirping of one or both of the lasers, such that 
\be
\lvert \omega_j ^{-1}\frac{d}{d t}\omega_j \rvert  \ll \omega_p \ll \omega_j
\ee
and
\be
 \lvert  \omega_1 -  \bar{\omega}_1 \rvert \lesssim  \lvert  \omega_2-  \bar{\omega}_2 \rvert \lesssim \lvert  \omega_1 -  \omega_2 \rvert 
 \sim \lvert  \bar{\omega}_1 -  \bar{\omega}_2 \rvert \sim \omega_{p},
 \ee
 where we define the average carrier frequencies
 \be
 \bar{\omega}_{j} = \frac{1}{T}\int\limits_{0}^{T} dt' \, \omega_j(t') = \left[ \psi_j(z = 0, t = 0) - \psi_j(z = 0, t = T)\right]/T,
 \ee
and the overall average carrier frequency is then defined as
 \be
 \bar{\omega} = \smallhalf \left(\bar{\omega}_1 + \bar{\omega}_2\right),
 \ee
Although taken as piecewise constant here, the amplitudes $a_j,\,  j = 1, 2$ could more generally also include suitably slow (\ie, slower than the plasma frequency, but possibly comparable to the chirp rate) time and position dependence as well, to model more realistic ramping of the drive fields and some 2D effects.
 
Within the plasma ($0 \le  z  \le  L$), we assume that the local laser frequencies $\omega_j(z, t) = -\frac{\del }{\del t} \psi_j$ and local wavenumbers $k_j(x, t) = \frac{\del }{\del z} \psi_j > 0$ each satisfy the 1D electromagnetic dispersion relation
\be\label{eqn:dispersion1}
\omega^2 = \frac{\omega_p^2}{\gamma_0} + c^2 k^2,
\ee
where the constant $\gamma_0 \ge 1$ parameterizes an average global nonlinear shift in the effective plasma frequency due to transverse electron quiver.  For electromagnetic waves satisfying this dispersion relation, the group velocity $v_g$ is given by
 \be\label{eqn:group_velocity}
v_g \equiv v_g(\omega) \equiv \frac{d}{d k}\omega(k) = c^2\frac{k(\omega)}{\omega} = c\left[ 1- \frac{\omega_p^2}{\gamma_0\omega^2} \right] ^{1/2},
\ee
which in our underdense case may be approximated as
\be
v_g(\omega) \approx  c\biggl[ 1- \smallhalf\frac{\omega_p^2}{\gamma_0\omega^2} + \ldots  \biggr].
\ee
We take as our reference group velocity $\bar{v}_g,$ the expression (\ref{eqn:group_velocity}) evaluated at the average carrier frequency $\bar{\omega}:$
\be
 \bar{v}_g \equiv  c\beta_p \equiv v_g(\bar{\omega})  \approx c\left[ 1- \smallhalf  \frac{\omega_p^2}{\gamma_0\bar{\omega}^2} \right]
\approx \frac{\bar{\omega}_2 - \bar{\omega}_1 }{k(\bar{\omega}_2) - k(\bar{\omega}_1)},
\ee
which represents the characteristic velocity at which both laser envelope modulations and Langmuir phase-fronts travel.

Because the relevant time-scales satisfy $\omega_{j}^{-1} \ll \omega_p^{-1} \ll T,$
 the fast carrier oscillations in $\abs{\boldsymbol{a}}^{2}$ at the harmonics $2\,\omega_1$ and $2\,\omega_2$ and the sum frequency $\omega_1 +  \omega_2$ will average away, leaving only the slowly-varying contribution to $\abs{\boldsymbol{a}}^{2}:$
\be\label{eqn:a_squared}
a^2 = \langle \abs{\boldsymbol{a}}^{2} \rangle =  \smallhalf\left[  \lvert a_1 \rvert^2 +  \lvert a_2 \rvert^2 
+a_1\cc a_2 e^{i\left( \psi_2 - \psi_1\right)}  + a_1 a_2\cc e^{-i\left( \psi_2 - \psi_1\right)} \right]. 
\ee
Since  the group-velocity dispersion effects ($\frac{d^2 }{d k^2}\omega$ and higher-order terms) remain small in the underdense regime, linear propagation into the plasma results in a beat phase given by
\be
\psi_2(z, t) - \psi_1(z,t)  = \Delta \psi_{0} + \int\limits_{0}^{t - z/\bar{v}_g}\!\!\!dt'\, \left[\omega_2(t')  - \omega_1(t')\right] \\
+ O\left(\frac{1}{\gamma_0}\frac{\omega_p^3}{\bar{\omega}^3} \frac{\omega_p L}{c}\right),
\label{eqn:beat_phase1}
\ee
where $\Delta\psi_{0}$ is just a constant, equal to the difference of the initial laser phases.
The neglected terms limit the validity of the constant group velocity approximation to interactions lengths $L$ less than the so-called dispersion length $L_{\text{\tiny{disp}}}:$
\be
L \le L_{\text{\tiny{disp}}} \sim \frac{\bar{\omega}^3}{\omega_p^3} \frac{c}{\omega_p}.
\ee
For accelerator applications,  the useful interaction length is already limited by the dephasing length $L_{\text{\tiny{d}}}$, beyond which accelerated electrons cannot gain energy from the electrostatic field:
\be
L \le L_{\text{\tiny{d}}} \sim  \frac{\bar{\omega}^2}{\omega_p^2} \frac{c}{\omega_p} \sim \frac{\omega_p}{\bar{\omega}}L_{\text{\tiny{disp}}} \ll L_{\text{\tiny{disp}}},
\ee
so that our constant group velocity approximation imposes no further restriction on the interaction length.

By appropriate choice of the initial phases, we may take both $a_1$ and $a_2$ to be real and nonnegative.  Defining the co-moving normalized beat frequency
\be
\Delta\omega(\xi) = \left[\omega_2(t = \xi/\omega_p) - \omega_1(t = \xi/\omega_p)\right]/\omega_p,
\ee
the beat phase
\be
\psi(\xi) = \psi(0) + \int\limits_{0}^{\xi} d\xi' \Delta\omega(\xi'),
\ee
for $\psi(0) \equiv \Delta\psi_{0},$ 
the normalized beat amplitude 
\be
\epsilon = a_1 a_2,
\ee
the average normalized intensity per laser
\be
\bar{a}^2 = \smallhalf\left[ a_1^2 + a_2^2 \right],
\ee
and the electron quiver factor
\be
\gamma_0 = \sqrt{1 + \bar{a}^2},
\ee
the reference group velocity $\bar{v}_{g}$ is fully determined, and 
the slow part of the normalized ponderomotive drive may be written as a function of $\xi$ only:
\be
a^2 = a^2(\xi)  \approx \Theta\left(\xi \right)\Theta\left(\omega_p T - \xi \right)\left[ \bar{a}^2 + \epsilon \cos\psi(\xi)\right],
\ee
where $\Theta(\xi)$ is the usual Heaviside step function.  Using this form for the ponderomotive drive, the equation of motion (\ref{eqn:quasi_simpler}) may be written as an ordinary differential equation in the co-moving coordinate $\xi:$
\be\label{eqn:quasi_eom}
\frac{d^{2}}{d \xi^{2}}\phi = \phi''(\xi) =  \frac{1}{2}\left[\frac{1 + \bar{a}^2 + \epsilon \cos\psi(\xi)}{(1+\phi)^{2}} -1\right].
\ee
with the initial conditions $\phi(\xi = 0) = \phi'(\xi = 0) = 0,$ valid
for $\tau > 0$ and $\tau- \omega_{p}L/\bar{v}_g \le \xi \le \tau,$ while otherwise $\phi(\xi, \tau) \equiv 0.$
Equation (\ref{eqn:quasi_eom}) is used for all numerical simulations discussed subsequently, and is the starting point for our analysis of autoresonance.

% section on: hamiltonian formalism

\section{Hamiltonian Formalism}\label{section:hamiltonian}

To study autoresonance, we now develop the Hamiltonian formulation of 
(\ref{eqn:quasi_eom}).  Our goal is an expression in terms of canonical action-angle 
variables, for which the phase-locking phenomenon is most readily 
analyzed.  First, note that the dynamical equation for the electrostatic 
potential (\ref{eqn:quasi_eom}) can be derived from the Hamiltonian
\be
  \mathcal{H}(\phi,p; \xi) = \frac{1}{2}p^2 + \frac{1}{2}\left[\frac{1}
	{1+\phi} + \phi - 1\right] + \frac{     \bar{a}^2 + \epsilon \cos\psi(\xi)    }{2(1+\phi)}	
 \equiv \mathcal{H}_0(\phi, p) + \frac{ \bar{a}^2 + \epsilon \cos\psi(\xi)}{2(1+\phi)},		\label{eqn:hamil}
\ee
with the scalar potential $\phi$ regarded as the generalized coordinate, $p \equiv \phi' \equiv \frac{d}{d\xi} \phi = \beta_p \varepsilon_{\|}$ regarded as  the canonical momentum conjugate to $\phi,$ and $\xi$ taken as the time-like evolution variable.  In this way, the plasma-wave dynamics are seen to be analogous to those of a one-dimensional forced nonlinear oscillator.  The component $\mathcal{H}_0$ of (\ref{eqn:hamil}) represents the Hamiltonian of the free oscillator, involving one term $\mathcal{T}(p) = \smallhalf p^2$ analogous to the kinetic energy of the oscillator (which is in fact proportional to the electrostatic {\it potential} energy density of the plasma wave), and another term corresponding to an anharmonic effective potential $\mathcal{V}(\phi) =  \smallhalf\left[1/(1+\phi) + \phi - 1\right].$  The remaining driving term $\Delta \mathcal{H} \equiv \mathcal{H}(\phi, p; \xi) - \mathcal{H}_0(\phi, p)$ corresponds to the time-dependent forcing of the oscillator.  This forcing is external in that it depends on a prescribed function of $\xi,$ but because of the nature of the relativistic nonlinearity, the effective strength of this forcing depends on both the intensity of the driving lasers and the instantaneous value of the dynamical variable $\phi$ associated with the excitation.  Because of the ponderomotive nature of the coupling, the effective forcing is always positive, \ie, includes a constant as well as a purely oscillatory part.  Both of these features differ somewhat from previously-studied autoresonant  systems, and will have important implications for the dynamics.

In the absence of forcing (\ie, $\epsilon = \bar{a}^2 = 0$), the dynamics governed by $\mathcal{H}_0(\phi, p) = \mathcal{T}( p) + \mathcal{V}(\phi)$ are conservative, so the oscillator energy, \ie, the value $H$ of the Hamiltonian $\mathcal{H}_0$ along any particular unperturbed orbit, remains constant.  In the physically allowed region $\phi > -1,$ the effective potential $\mathcal{V}(\phi)$ is nonnegative and possesses a single minimum $\mathcal{V} = 0$ at $\phi = 0,$ while $\mathcal{V}(\phi) \to +\infty$ as $\phi \to -1^{+}$ or $\phi \to +\infty.$  So for any value of energy $H$ satisfying $0 \le H < \infty,$ there exists a phase space trajectory $\bigl(\phi(\xi; H), \phi'(\xi; H) \bigr)$ which is a closed periodic orbit, mirror-symmetric about its turning points, for which $-1 < \phi(\xi) < \infty$ throughout, and which is unique up to its overall phase, or initial position at $\xi = 0.$

Making a canonical transformation to the action-angle variables 
of the free oscillator, $\phi = \phi(\mathcal{I}, \theta)$, $p = 
p(\mathcal{I}, \theta)$, we can express (\ref{eqn:hamil}) as
\begin{equation}
  \mathcal{H}(\mathcal{I}, \theta; \xi) = \mathcal{H}_0(\mathcal{I}) + \frac{  \bar{a}^2 + \epsilon \cos\psi(\xi)  }{2[1+\phi(\mathcal{I},\theta)]},
						\label{eqn:act_hamil}
\end{equation}
where the action $\mathcal{I}$ is defined in terms the area in phase space contained within the unperturbed orbit $\left(\phi(\xi; H), \phi'(\xi; H) \right)$ of energy $H$:
\begin{equation}
  \mathcal{I} \equiv \frac{1}{2\pi}\oint p\,d\phi =  \frac{1}{\pi}
	\int\limits_{\phi_-}^{\phi_+}\phi'\,d\phi.  \label{eqn:action}
\end{equation}
Here, $\phi_+$ and $\phi_-$ are, respectively, the upper and lower turning points of the orbit, at which $\mathcal{V}(\phi_{\pm}) = H:$
\begin{equation}
  \phi_\pm = H \pm \sqrt{H^2 + 2H},
\end{equation}
and in (\ref{eqn:action}) we have used symmetry to reduce the integration path to the segment where $p \ge 0.$   By making the change of variables $\phi = \phi_+ - (\phi_+ - \phi_-)\sin^2 (u)$, the action (\ref{eqn:action}) can be calculated with the help 
of a standard integral table, \eg, equation 2.584-13 on p. 163 of \cite{gradshteyn_ryzhik:80}:
\begin{eqnarray}
  \mathcal{I} &=& \frac{2(\phi_+ - \phi_-)^2}{\pi\sqrt{1 + \phi_+}} \int 
	\limits_0^{\pi/2} d u\, \frac{\sin^2(u) \cos^2(u)}
	{\sqrt{1 - \frac{\phi_+ - \phi_-}{1 + \phi_+}\sin^2(u)}}
						\nonumber \\
  &=& \frac{4}{3\pi} \left[1 + H - \sqrt{H^2 + 2H} \right]^{1/2}
	\left\{\frac{(1 + H)E(\kappa)}{1 + H - \sqrt{H^2 + 2H}} - K(\kappa)\right\}.
						\label{eqn:i_h}
\end{eqnarray}
Here, $K(\kappa)$, $E(\kappa)$ are complete elliptic integrals of the first and 
second kind, respectively, whose modulus satisfies
\begin{equation}
 \kappa = +\left[ 2(1+H)\sqrt{H^2+2H} - 2H(2+H)\right]^{1/2}.  
\end{equation}
At this point, one could in principle use (\ref{eqn:i_h}) to find 
$\mathcal{H}_0(\mathcal{I})$, but fortunately such a cumbersome inversion will not 
be necessary.  The normalized (\ie, dimensionless) nonlinear frequency $ \Omega(H)$ of the unforced oscillator is given by
\begin{equation}
  \Omega(H) \equiv \frac{d}{d \xi}\theta = \frac{\partial \mathcal{H}_0} 
	{\partial \mathcal{I}} = \left(\frac{\partial \mathcal{I}}
	{\partial H} \right)^{-1} = \frac{\pi}{2}\frac{\left[1 + H - \sqrt{H^2+2H} 
	\right]^{1/2}} {E(\kappa)}.			\label{eqn:non_lin_w}
\end{equation}
To put (\ref{eqn:hamil}) in the desired form (\ref{eqn:act_hamil}), the 
remaining ingredient needed is the canonical transformation $\phi = 
\phi(\mathcal{I},\theta)$.  An explicit formulation of this will also not 
be needed, and we may proceed by formally assuming that we have made this 
substitution.  Then, $\phi = \phi(\mathcal{I}, \theta)$ is a periodic 
function of $\theta$, and we can expand the driving term appearing in
(\ref{eqn:act_hamil}) in a Fourier series as
\begin{equation}
 \frac{ \bar{a}^2 + \epsilon \cos\psi(\xi) }{2\left[1+ \phi(\mathcal{I}, \theta)\right]} = 
\left[ \bar{a}^2 + \epsilon \cos\psi(\xi)\right]
\sum_{n=-\infty}^{\infty}b_n(\mathcal{I})e^{in\theta}.
						\label{eqn:phi_sum}
\end{equation}
Because (\ref{eqn:phi_sum}) is a real-valued  function, the Fourier coefficients 
must satisfy $b_{n} = b_{-n}^{*}$, and are defined by
\begin{equation}
  b_{n}(\mathcal{I}) = \frac{1}{2\pi}\int\limits_{0}^{2\pi}d\theta\, 
	\frac{e^{-in\theta}}{2[1 + \phi(\mathcal{I},\theta)]}.
						\label{eqn:four_coef}
\end{equation}
We put (\ref{eqn:four_coef}) into a form more amenable to calculation by 
changing variables using $\theta = \Omega(H)\xi$, and integrating over one 
period $0 \le \xi \le  \Xi$ of the co-moving coordinate, where $\Xi \equiv \Xi(H) \equiv 2\pi/ \Omega(H).$  Furthermore, if we choose the origin of the orbit associated with the energy $H,$ or equivalently with the free action $\mathcal{I}(H),$ such that $\phi(\xi = 0; H) = \phi_+$, the potential $\phi(\xi; H),$ and hence $\left[1+\phi(\xi; H)\right]^{-1},$ are symmetric 
about the point $\xi = \Xi/2.$  Since the imaginary parts of (\ref{eqn:four_coef}) are 
obtained by integrating over the antisymmetric functions $\sin(n\Omega \xi)$, the 
$b_n$'s are purely real and given by:
\begin{equation}
  b_n(H) =  b_n(H)\cc = b_{-n}(H) = b_{-n}(H)\cc = \frac{1}{\Xi(H)}\int\limits_0^{\Xi(H)} d\xi\, 
	\frac{\cos\left[n\Omega(H)\xi\right]}{2[1 + \phi(\xi; H)]}.
\end{equation}

Up to this point, no approximations have been made in the canonical transformations to the action-angle variables of the unforced oscillator.   Now, we make the Single Resonance Approximation (SRA) \cite{chirikov:79} to 
(\ref{eqn:act_hamil}), by assuming that the rapidly oscillating terms of 
the Hamiltonian average to zero and contribute negligibly to the 
dynamics.  Anticipating the developments of Section \ref{section:autoresonance}, we realize that 
under certain frequency-locking conditions derived there, autoresonant excitation occurs,
wherein the plasma wave amplitude, or equivalently free-oscillator energy $H(\xi) = \mathcal{T}\left(\phi'(\xi)\right) + \mathcal{V}\left(\phi(\xi)\right),$ grows secularly such that  the 
dynamical frequency $\Omega(\xi)$ of the forced oscillator approximately matches that of the unforced oscillator at that same amplitude, $\Omega(\xi) \approx \Omega\bigl(H(\xi)\bigr).$ Simultaneously, this instantaneous oscillator frequency follows that of the driving beat frequency of the wave, $\Omega(\xi) \approx \Delta\omega(\xi).$ 
 Under these assumptions, we can consistently neglect all terms in the sum 
(\ref{eqn:phi_sum}) except the constant term or those terms with frequency dependence $\sim \pm\left[\Omega(\xi) - \Delta\omega(\xi)\right].$ The averaged Hamiltonian then becomes
\begin{equation}
  \mathcal{H}(\mathcal{I},\theta; \xi) = \mathcal{H}_0(\mathcal{I}) + 
	\epsilon b_1(\mathcal{I})\cos[\theta - \psi(\xi)] + 
	\bar{a}^2 b_0(\mathcal{I}).
\end{equation}

We now make an explicitly $\xi$-dependent canonical transformation to the
rotating action-angle variables $(\hat{\mathcal{I}},\Psi)$.  Using the 
mixed-variable generating function
\be
F_2(\hat{\mathcal{I}},\theta;\xi) = \left[\theta - \psi(\xi)\right]
	\hat{\mathcal{I}},
\ee
our old and new coordinates are related by
\begin{subequations}
\begin{align}
 \Psi &= \frac{\partial F_2}{\partial\hat{\mathcal{I}}} = \theta - \psi(\xi);\\
 \mathcal{I} &= \frac{\partial F_2}{\partial\theta} = \hat{\mathcal{I}}.
\end{align}
\end{subequations}
Dropping carets from the new action, the Hamiltonian in the rotating frame is
\begin{equation}
  \mathcal{K}(\mathcal{I},\Psi;\xi) =
	\mathcal{K}(\mathcal{I},\theta,\xi) +
	\frac{\partial F_2}{\partial\xi} \nonumber 
  = \mathcal{H}_0(\mathcal{I}) - \Delta\omega(\xi)\mathcal{I} +
	\epsilon b_1(\mathcal{I}) \cos\Psi + \bar{a}^2b_0(\mathcal{I}).
					\label{eqn:sing_res}
\end{equation}
The resulting  SRA canonical equations of motion are
\begin{subequations}\label{eqn:sra_eom2}
\begin{align}
  \frac{d}{d\xi}\Psi &= \phantom{-}\frac{\del \mathcal{K}}{\del \mathcal{I}}
 = \Omega(\mathcal{I}) - \Delta\omega(\xi) + 
	\epsilon\frac{\partial b_1}{\partial\mathcal{I}}\cos\Psi +
	\bar{a}^2\frac{\partial b_0}{\partial\mathcal{I}}, 
						\label{eqn:psi_dot} \\
\frac{d}{d\xi}\mathcal{I} &= -\frac{\del \mathcal{K}}{\del\Psi} = \epsilon b_1(\mathcal{I})\sin\Psi,  \label{eqn:i_dot}
\end{align}
\end{subequations}\label{eqn:sra_dot}
for which we next determine the required conditions for phase-locking.

% section on: autoresonant response
 
\section{Autoresonant Response}\label{section:autoresonance}

The essential ingredients for autoresonance, as exploited in the present scheme, are: a nonlinear, oscillatory degree of freedom (in our case, the plasma wave) evolving, in the absence of any forcing, within an integrable region of phase space;  a continuous functional relationship between the nonlinear frequency and energy of the oscillation,  possessing a well-defined linear limit; an applied oscillatory driving force (in our case, the modulated ponderomotive envelope of the lasers) which is sufficiently small as to be considered perturbative, so that the notion of the nonlinear frequency of the unforced oscillator remains meaningful; and initial conditions and forcing profile consistent with Adiabatic Passage Through Resonance (APTR) --  namely, an initially unexcited system (quiescent plasma), and an initial drive frequency sufficiently far from the linear resonance, with subsequent time-dependence that is sufficiently slowly-varying but otherwise arbitrary.

The key consequence of the autoresonant beat-wave generation is the robust entrainment 
between the three relevant (normalized) frequencies: the beat frequency $\Delta\omega(\xi)$ of the driving lasers, the instantaneous frequency $\Omega(\xi)$ of the driven plasma wave, and the nonlinear frequency $\Omega(H)$ of the unforced plasma wave.  Assuming this phase-locking is achieved, amplitude control of the plasma wave can be simply 
understood via (\ref{eqn:non_lin_w}): because the frequency of the 
freely-evolving nonlinear plasma wave is a function of the energy, phase-locking of the driven wave to the envelope such that  $\Delta\omega(\xi) \approx \Omega(\xi) \approx \Omega(H)$ implies that changing the drive frequency will correspondingly change the oscillator energy.  In our case, 
(\ref{eqn:non_lin_w}) indicates that $\Omega(H)$ is a decreasing function 
of the energy, so that in order to increase the plasma wave amplitude one 
must decrease the beat frequency as a function of the co-moving position $\xi$ (or, equivalently,  in time $t$ at the source of the driving lasers.) 

While we have indicated how autoresonance can lead to large plasma waves, 
we have not yet shown under what conditions such phase-locking occurs.  To 
answer this question, we first consider the linear and weakly nonlinear 
response, valid up to the point where the normalized sweeping drive frequency is of the order of the 
normalized linear resonance, $\Omega(H) \approx \Delta\omega(\xi) \gtrsim 1$.  Next, we consider the fully nonlinear case, for which we derive adiabaticity requirements for autoresonance.

\subsection{Small Amplitude Response and Phase-Locking}

When the drive is first applied with its frequency above the linear 
resonance, the plasma wave amplitude (and, hence, $H$ and $\mathcal{I}(H)$) 
are small and we can linearize equations 
(\ref{eqn:psi_dot} - \ref{eqn:i_dot}).  In this limit, the oscillator is harmonic, with $\Omega(H) = 1$, $\phi(\xi) = \sqrt{2H}\cos(\xi)$, and $H = \mathcal{I}$, so that $b_0 = -\smallhalf H$, $b_1 = \sqrt{2H}/4$, and equations
(\ref{eqn:sra_eom2}) become
\begin{subequations}\label{eqn:lin_eom2}
\begin{align}
 \frac{d}{d\xi}\Psi &= 1 - \smallhalf\bar{a}^2 - \Delta\omega(\xi) + 
	\tfrac{\epsilon}{4\sqrt{2}}\frac{1}{\sqrt{\mathcal{I}}}\cos\Psi \label{eqn:lin_psi}, \\
  \frac{d}{d\xi}\mathcal{I} &= \tfrac{\epsilon}{2\sqrt{2}}
	 \sqrt{\mathcal{I}} \sin\Psi. 		\label{eqn:lin_i}					
\end{align}
\end{subequations}\label{eqn:lin_eom}
Note that the $(1 - \half\bar{a}^2)$ contribution to  $\frac{d}{d\xi}\Psi$ corresponds, in normalized units, to the leading order expansion of the effective plasma frequency in the EM dispersion relation:
\be
\omega_{p_{\text{\tiny{eff}}}} \equiv \omega_p/\gamma_0 = \omega_p/\sqrt{1 + \bar{a}^2} \approx \omega_p\left(1 - \smallhalf\bar{a}^2 + \ldots \right),
\ee
which is shifted from the bare value $\omega_p$ due to the transverse quiver motion of the electrons in the applied laser fields.  While we never explicitly invoked any small $a^2$ approximation, only the leading-order correction appears because, in making the SRA, we ignored terms of the form $\bar{a}^{2}e^{ i n \theta }$ for $\lvert n \rvert \ge 1.$  But for sufficiently large intensity $\bar{a}^2$ such terms can appreciably effect the motion despite being off-resonance.  One could, \textit{a posteriori}, partially correct for this defect by replacing $(1 - \smallhalf \bar{a}^2)$ with $\left(1 + \bar{a}^2 \right)^{-1/2}$ in the equation of motion (\ref{eqn:lin_psi}), but this will be unnecessary for the small values of  $\bar{a}$ considered here.  Physically, we should simply ensure that the drive frequency begins sufficiently far above, and then is slowly swept past, the effective (or renormalized) frequency $\omega_{p_{\text{\tiny{eff}}}}$, rather than the bare frequency $\omega_p,$ if the difference is not negligible.  Precise knowledge of the exact value of the effective linear plasma resonance including the effects of transverse quiver is not needed.  

For the driven plasma wave, we have $\Delta\omega(\xi) \sim \Omega(\xi) \sim 1,$ while we seek solutions for which $\lvert \Delta\omega(\xi)- \Omega(\xi) \rvert \ll 1$ as a result of special initial and forcing conditions: an initially unperturbed plasma, $\phi(\xi = 0) = \phi'(\xi = 0) = 0;$ an initial tuning of the beat frequency above resonance, \ie,  $\Delta\omega(\xi = 0) > \omega_{p_{\text{eff}}},$ and subsequently a slow downward frequency chirp through resonance, where the chirp rate is be characterized by a parameter $\alpha \equiv \alpha(\xi) \equiv -\frac{d}{d\xi}\Delta\omega(\xi),$ with $0 \le \alpha \ll 1.$

For a linear  frequency chirp around the effective plasma frequency,
\be\label{eqn:linear_chirp}
\Delta\omega(\xi) = 1 - \smallhalf \bar{a}^2 - \alpha\xi,
\ee
the simple harmonic oscillator equations (\ref{eqn:lin_eom2}) have analytic solutions in terms of the Fresnel Sine and Cosine integrals \cite{lewis:32}.  We briefly summarize the extensive characterization of these solutions found in \cite{friedland:92}.  When the drive is first applied far from resonance, the oscillator response can be divided into two components: one 
ringing component precisely at $\omega_{p_{\text{\tiny{eff}}}}$, and the other at the driving 
frequency $\Delta\omega(0),$ both of small amplitude.  The singular term 
$\sim\mathcal{I}^{-1/2}$ in (\ref{eqn:lin_psi}) allows for a large change in 
phase at small amplitude without violating the requirement that $\phi(\xi)$ remain smooth, so that the response at the driven frequency can adjust itself to the drive, and phase-locking can occur.  As the frequency is swept toward the resonance, these driven, phase-locked oscillations grow.  
Meanwhile, the response at the resonant frequency has no such phase relation 
with the drive, and remains small.  In this way, we essentially have one 
growing, phase-locked plasma wave when the drive frequency reaches the resonant frequency.

\begin{figure}
  \centering
  \includegraphics[scale=1]{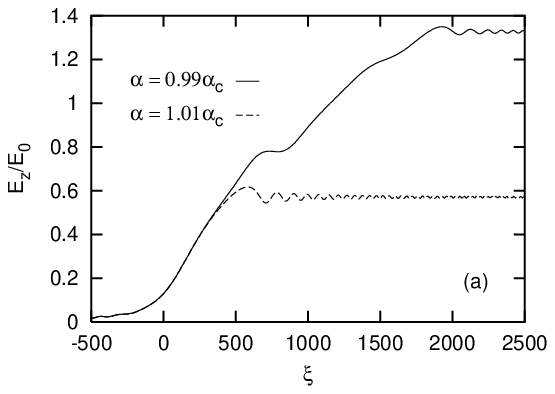}
  \includegraphics[scale=1]{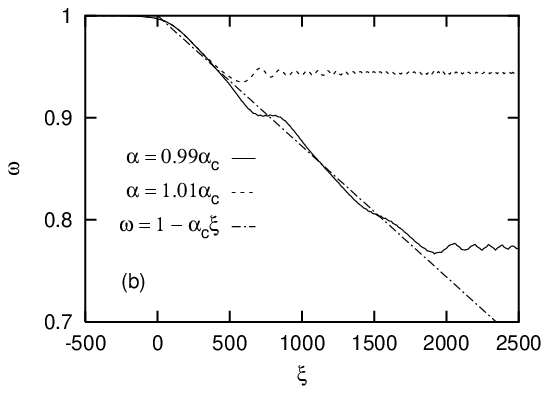}
  \caption{Demonstration of the critical autoresonant behavior for 
	$\epsilon = 0.005$.  In (a), chirp rates 1\% below 
	critical (solid line) yield final longitudinal fields above linear 
	wavebreaking, whereas chirp rates 1\% above saturate at $0.6E_0$.  
	(b) compares the phase locking in autoresonance (solid 
	line) to that of non-autoresonant behavior (dashed line).  The 
	critical drive (mixed line) is included for reference.}
						\label{fig:ex_005}
\end{figure}

As the resonance is approached, the amplitude of the plasma wave 
begins to become large and one must account for the growing nonlinearities.  
We therefore expand the expression for the free nonlinear frequency (\ref{eqn:non_lin_w}) to first 
order:
\begin{equation}
  \Omega(H) = 1 - \tfrac{3}{8}H = 1- \tfrac{3}{8}\mathcal{I},
\end{equation}
and continue to use the linearized frequency chirp (\ref{eqn:linear_chirp}).
Making the change of variable $\mathcal{A} \equiv 
4\sqrt{2\mathcal{I}}$, we have
\begin{subequations}
\begin{align}
\frac{d}{d\xi}\Psi &= \alpha\xi - \tfrac{3}{256}\mathcal{A}^2 + 
	\frac{\epsilon}{\mathcal{A}}\cos\Psi, \\
  \frac{d}{d\xi} \mathcal{A} &= \epsilon \sin\Psi.
\end{align}
\end{subequations}
This set of equations can be reduced to a single first-order ordinary 
differential equation by defining the complex dynamical variable $Z \equiv 
-\sqrt{256/3\alpha}\mathcal{A}e^{i\Psi}$, a re-scaled independent variable $\zeta \equiv 
\sqrt{\alpha}\xi,$ and the dimensionless parameter $\mu \equiv 
\epsilon\sqrt{3/(256\alpha^{3/2})}$, to obtain
\begin{equation}
  i\frac{d}{d\zeta}Z + \left(\zeta - \lvert Z \rvert^2 \right)Z = \mu.
						\label{eqn:NLS}
\end{equation}
Thus, the weakly nonlinear problem is now described by a dynamical equation with a 
single parameter $\mu$ that combines the drive strength $\epsilon$ and the chirp 
rate $\alpha.$  It has been found numerically \cite{grosfeld_friedland:02} that the 
solution to (\ref{eqn:NLS}) has a bifurcation at the critical value $\mu = 
\mu_c \approx 0.411.$  For $\mu < \mu_c$, the plasma wave 
response quickly dephases from the drive, resulting in only small excitations.  
In contrast, for $\mu > \mu_c,$ phase-locking occurs and the plasma 
wave can grow to large amplitude.  This critical behavior with respect to $\mu$ 
translates into a critical drive strength $\epsilon$ and chirp rate 
$\alpha$ for the nonlinear oscillator to be autoresonantly excited.   The 
condition is
\begin{equation}
  \alpha \leq \left(\frac{3\epsilon^2}{256\mu_c^2}\right)^{2/3} \approx 0.169\,\epsilon^{4/3}.			\label{eqn:alpha_crit}
\end{equation}
Thus, for a given laser intensity one can readily find the maximum chirp rate that can be tolerated and still 
obtain high amplitude plasma waves.  We demonstrate 
the sensitivity of this critical behavior for $\epsilon = 0.005$ in 
Fig.~\ref{fig:ex_005}, which plots numerical solutions to the full quasi-static equation of motion (\ref{eqn:quasi_eom}).  Plot (a) shows the excited longitudinal electric field for chirp rates just above and just below the critical rate $\alpha_c \approx 1.28 \times 10^{-4}$; Plot (b) demonstrates the dynamic 
frequency-locking that occurs in autoresonance, and compares this to the 
case where autoresonance fails to occur.

\subsection{Fully Nonlinear Autoresonant Response}

If phase-locking is maintained through the weakly nonlinear regime, the 
amplitude continues to grow and one must consider further nonlinearities beyond those included in (\ref{eqn:NLS}).  In this case, there arises more stringent restrictions on the chirp rate for adiabatic phase-locking to persist.  We calculate this condition by first finding a second-order equation for the phase $\Psi$:
\begin{equation}\label{eqn:psi_nl}
  \frac{d^2}{d\xi^2}\Psi = \{\frac{d\Psi}{d\xi},\mathcal{K} \} + 
	\frac{\partial}{\partial\xi}\frac{d\Psi}{d\xi} 
  = \frac{\partial^2\mathcal{K}}{\partial\mathcal{I}\partial\Psi}
	\frac{\partial\mathcal{K}}{\partial\mathcal{I}} - 
	\frac{\partial^2\mathcal{K}}{\partial\mathcal{I}^2} 
	\frac{\partial\mathcal{K}}{\partial\mathcal{I}} + 
	\frac{\partial^2\mathcal{K}}{\partial\mathcal{I}\partial\xi} +
	\frac{\partial}{\partial\xi}\frac{d\Psi}{d\xi},
\end{equation}
where we have made use of the usual canonical Poisson bracket with respect to the rotating-frame action-angle variables:
\be
\{F\,,G\} \equiv \frac{\partial F}{\partial\Psi}
\frac{\partial G}{\partial\mathcal{I}} - \frac{\partial F}{\partial\mathcal{I}}\frac{\partial G}{\partial\Psi}.
\ee

Using this expression (\ref{eqn:psi_nl}) as well as  (\ref{eqn:psi_dot}), the phase $\Psi$ is seen to obey the following second-order equation:
\begin{equation}
  \frac{d^2}{d\xi^2}\Psi + 
	\epsilon\frac{\partial b_1}{\partial\mathcal{I}}\sin\Psi 
	\frac{d}{d\xi}\Psi
	 + \left\{\tfrac{d}{d\xi}\Delta\omega - \epsilon b_1(\mathcal{I})\sin\Psi
	\left[\frac{\partial\Omega}{\partial\mathcal{I}} + 
	\bar{a}^2\frac{\partial^2 b_0}{\partial\mathcal{I}^2} + 
	\epsilon\frac{\partial^2 b_1}
	{\partial\mathcal{I}^2}\cos\Psi\right]\right\} = 0.
						\label{eqn:psi_ddot}
\end{equation}
Now, we assume (see, \eg, \cite{fajans_et_al:99}) that the free action can be written as 
\be\label{eqn:action_form}
\mathcal{I} = \mathcal{I}_0 +  \Delta\mathcal{I},
\ee
where $\mathcal{I}_0 = \mathcal{I}_0(\xi)$ is the slowly-varying, secularly-evolving action about which there 
are small oscillations given by $\Delta\mathcal{I} = \Delta\mathcal{I}(\xi)$.  These oscillations 
correspond to fluctuations in $\Psi$ about its (slowly-varying) phase-locked value $\bar{\Psi} =\bar{\Psi}(\xi),$ an example of which can be seen in Fig.~\ref{fig:ex_005}(b).  In the autoresonant case (solid line), we see that as the plasma wave is excited, its frequency does indeed make small 
oscillations about the drive frequency.  We further note that as one decreases the chirp rate $\alpha$ from its critical value $\alpha_c$, we obtain more total oscillations in frequency over the longer excitation time, but with a slightly smaller magnitude of excursions from the drive frequency.  

Using the form (\ref{eqn:action_form}) for the 
action, the lowest-order equation for the phase is identical to 
(\ref{eqn:psi_ddot}), with $\mathcal{I}$ being replaced everywhere by 
$\mathcal{I}_0$.  In this way, the phase itself is seen to obey a nonlinear oscillator 
equation, with an effective nonlinear damping (or anti-damping) term, and a conservative ``forcing,''  described  by the terms in the braces, which is derivable from an effective ``potential'' (not to be confused with any previously-mentioned potential) whose shape is dictated by the slowly-evolving action $\mathcal{I}_0$ and the drive parameters $\epsilon$ and $\alpha.$  Phase-locking then corresponds to trapping of $\Psi$ in a basin of this effective potential.  In order for the phase $\Psi$ to be trapped about some value $\bar{\Psi},$ the effective potential must possess a local minimum there, and the non-conservative term must either provide damping or else remain sufficiently small if excitatory.  In fact, the non-conservative term is expected to be small compared to the second term in the conservative force. Their ratio is given by $\left(\tfrac{1}{b_1}\tfrac{\partial b_1}{\partial\mathcal{I}_0}/ \tfrac{\partial\Omega}{\partial\mathcal{I}_0}\right) \left(\tfrac{d \Psi}{d \xi} \right),$ where typically 
$\left(\tfrac{1}{b_1}\tfrac{\partial b_1}{\partial\mathcal{I}_0}/ \tfrac{\partial\Omega}{\partial\mathcal{I}_0}\right) \sim 
\left(\tfrac{1}{b_1}\tfrac{\partial b_1}{\partial\Omega} \right) \sim \frac{1}{\Omega} \sim O(1)$, while 
$\tfrac{d \Psi}{d \xi} \ll O(1)$ because the nonlinear phase oscillations are slow compared to $\omega_p.$  Actually, over most of the typical range of parameter values, the last two terms in the conservative forcing are small compared to the first two terms, since they are higher order in the drive strength, and as a first approximation the phase oscillations are governed by the ``biased'' pendulum equation
\be \label{eqn:psi_simpler}
 \frac{d^2}{d\xi^2}\Psi \approx \alpha(\xi) + \epsilon b_1(\mathcal{I}_{0})\frac{\partial\Omega}{\partial\mathcal{I}_0} \sin\Psi,
\ee
although we will continue to work with the full equation (\ref{eqn:psi_ddot}).  Clearly, for a given $\mathcal{I}_0,$ if the normalized chirp rate $\alpha$ remains sufficiently small compared to the normalized drive strength $\epsilon,$ then the effective potential will be of the tilted-washboard variety, with a series of periodically-spaced local minima in $\Psi$ at intervals of $2\pi.$   As $\alpha$ increases, the depth of these wells decreases, until they finally disappear, as does any opportunity for phase-locking.

Thus, a necessary condition for trapping is that the effective force can actually vanish at some fixed point $\bar{\Psi}$:
\be
 \alpha(\xi) + \epsilon b_1(\mathcal{I}_0)\sin\bar{\Psi}\left[
	\frac{\partial\Omega}{\partial\mathcal{I}_0} + 
	\bar{a}^2\frac{\partial^2 b_0}{\partial\mathcal{I}_0^2} + \epsilon
	\frac{\partial^2 b_1}{\partial\mathcal{I}_0^2}\cos\bar{\Psi}\right] = 0.
						\label{eqn:zero}
\ee
For given $\mathcal{I}_0,$ $\epsilon,$ and $\alpha,$ this determines the average, slowly-varying phase $\bar{\Psi}$ about which trapping occurs.  In the linear ($\mathcal{I}_0 \to 0$), weakly forced ($0 \le \epsilon \lll 1$) or weakly chirped ($\alpha \to 0^{+}$) regimes, this phase value is known \cite{friedland:92} to be $\bar{\Psi} \approx \pi,$ but for nonlinear plasma waves, stronger forcing, or faster chirping, this phase can shift appreciably.  Since $\lvert \sin\bar{\Psi} \rvert,\, \lvert \cos\bar{\Psi} \rvert  \le 1,$ equation (\ref{eqn:zero}) also imposes an upper bound on the frequency chirp $\alpha(\xi)$ for which the phase can remain trapped in autoresonance, regardless of the actual value of the phase at which locking occurs.  Setting $\lvert \sin\bar{\Psi} \rvert =  \lvert \cos\bar{\Psi} \rvert  = 1$ above, we obtain an upper bound on $\alpha$ beyond which any phase-locking is impossible:
\begin{equation}
 0 \le  \alpha(\xi) \le \epsilon \lvert b_1(\mathcal{I}_0) \rvert
	\left[\left\lvert \frac{\partial\Omega}{\partial\mathcal{I}_0}\right\rvert + 
	\bar{a}^2\left\lvert \frac{\partial^2 b_0}{\partial\mathcal{I}_0^2}\right\rvert  + 
	\epsilon \left\lvert \frac{\partial^2 b_1}{\partial\mathcal{I}_0^2}\right\rvert \right].
						\label{eqn:limit}
\end{equation}

\begin{figure}
  \centering
  \includegraphics[scale=1]{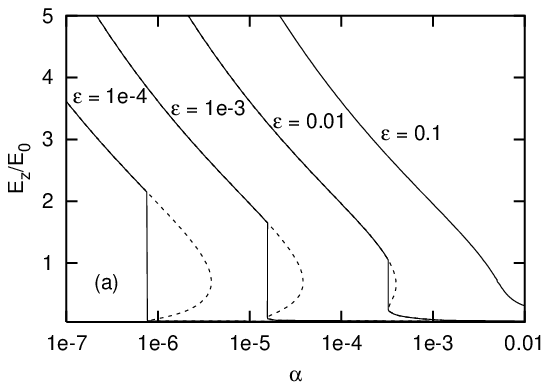}
  \includegraphics[scale=1]{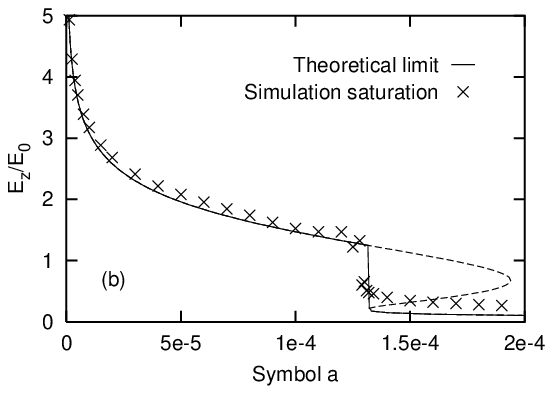}
  \caption{Shows the maximum plasma wave amplitude obtainable before the 
	``slowness'' condition (\ref{eqn:limit}) is violated.  In (a), we 
	plot the maximum longitudinal field as a function of the 
	chirp rate for four different driving strengths.  Dotted lines 
	indicate hysteresis at $\alpha_c$ given by (\ref{eqn:alpha_crit}).  
	(b) compares the theory (line) with numerically 
	determined saturation for $\epsilon=0.005$}
						\label{fig:chirp_E}
\end{figure}

Again, for realistic parameters, the force balance typically resides predominately between the first two terms in (\ref{eqn:zero}), and hence this upper bound, although approximate, is expected to provide a reasonably tight cutoff for autoresonant phase-locking, which has been confirmed by numerical simulation.   This inequality can also be thought of as giving the maximum achievable plasma wave amplitude (implicitly as a function of $\mathcal{I}_0$) for a given chirp rate and laser power.  We show the dependence of the saturated 
longitudinal field on the chirp rate as a solid line in 
Fig.~\ref{fig:chirp_E}(a) for a number of different drive strengths.  For a 
fixed drive strength $\epsilon$, the maximum attainable electric field jumps 
discontinuously at the critical chirp rate $\alpha_c$ given by 
(\ref{eqn:alpha_crit}).  The dotted lines correspond to solutions of 
(\ref{eqn:limit}) that cannot be accessed when starting from vanishing initial 
longitudinal field, and as such constitutes a form of hysteresis in 
the excitation.  The solid lines show the stable branches for the case of interest: excitation from
 a quiescent plasma via slow downward chirping from a drive initially above resonance. Fig.~\ref{fig:chirp_E}(b)  shows a comparison of the theoretical maximum amplitude and that found by numerically integrating the 
quasi-static equation of motion (\ref{eqn:quasi_eom}).

Asymptotic expansions and numerical plots for $\Omega(\mathcal{I})$ and $b_{n}(\mathcal{I})$ reveal that the right-hand-side of (\ref{eqn:limit}) actually decreases with increasing $\mathcal{I}_0$ (at least beyond moderate values of $\mathcal{I}_0$), making the adiabatic nonlinear phase-locking condition (\ref{eqn:limit}) become increasingly stringent as the plasma wave grows.  For fixed chirp rate and drive amplitude, the growing plasma wave will eventually fall out of frequency-locking.  In the physical pendulum limit discussed above, note that when this cutoff is reached and autoresonance is lost, the average phase should be near $\pm \pi/2.$  Mathematically, this saturation can be traced back to the nonlinear nature of the forcing in (\ref{eqn:quasi_eom}), where the effective strength of the forcing depends both on the external drive intensity and the plasma wave amplitude; physically, to the fact that as the plasma wave grows, the electrons become increasingly relativistic and therefore less susceptible to displacement via the ponderomotive forcing at a fixed intensity, so the effective strength of a given drive actually decreases.  In principle, this may be counteracted by either slowly increasing the laser amplitude or, more practically, by slowly decreasing the chirp rate in time.  In the latter case, the rate of growth of the plasma wave will fall off because it is directly tied to the decreasing drive frequency, but the wave can remain frequency-locked with the drive indefinitely (with some oscillating excursions.)

As a result of relativistic detuning, recall that in the original RL/TD scheme, the plasma wave amplitude exhibits slow (compared to $\omega_p^{-1}$) nonlinear modulations which appear as beating, \ie,  periodic amplitude oscillations up to the RL limit and back to a nearly unexcited state.  As the wave is driven to a high amplitude its phase slowly slips until it is more than $\sim \pi/2$ out of phase with respect to the laser beat and then gives its energy back to the lasers, then continues to shift further out of phase, only to be re-excited when its amplitude approaches zero and it can re-establish phase matching with the drive.  Both the frequency difference between plasma wave and drive and the phase-lag exhibit continuous oscillations in time.

In the DMG scheme, which relies on autoresonance but starts at linear resonance, the plasma wave amplitude not only can peak at a higher maximum than in the original scheme,  but typically will sustain a higher average value at long times, exhibiting a nonlinear ringing about some non-zero saturated value rather than a full  beating.  We suspect that this behavior results from two features of the chirping: when the autoresonant adiabatic condition fails, the wave phase is closer to its neutral value, with respect to energy exchange with the drive, and after autoresonance is lost, the frequency difference and phase lag each grow secularly in time rather than exhibiting oscillations.   Depending on initial parameters, the initial growth up to the absolute maximum can either be essentially monotonic, or exhibit a ``staircase'' behavior, with intermittent plateaus or dips at local maxima between periods of resumed growth, before finally leveling off with some ringing. When starting from resonance,  the originally published numerical examples \cite{deutsch_et_al:91} and our own simulations suggest that the observed ripples in the amplitude excitation are actually minimized at some intermediate value of the chirp rate, implying that the excitation may not be fully autoresonant in some sense.

In our autoresonant scheme based on APTR, the behavior more closely resembles that in DMG scheme, but exhibits more nearly monotonic growth, higher peak fields, and less ringing after saturation.  Specifically, some slow oscillations in the amplitude may be present early on, but once nonlinear phase-locking has been achieved, the growth remains monotonic or virtually so until finally the amplitude levels off and appears to almost saturate, with only a very small amount of subsequent ringing.

The extent of the amplitude and phase excursions, which affect  how regularly the excitation grows and how closely the phase is locked during autoresonant excitation, and also how much ringing persists after saturation, are determined by how deeply the phase is trapped in its effective potential well.  This in turn depends both on how steep and how deep is the available well (determined by the plasma wave amplitude and drive lasers parameters),  and to what extent the phase can be nudged into position near the bottom of the well and kept there (determined by the initial conditions and the adiabaticity of  the chirped forcing).

Numerical simulations suggest that, as one would expect, the extent or depth of  this trapping is improved by using a stronger drive (at least up to some moderately strong value), starting the drive frequency  further above resonance, and chirping more slowly.  In practice, of course, each of these strategies involves trade-offs.  Increasing the drive strength increases the growth rates for laser-plasma  instabilities that might disrupt the forcing.  Either increasing the initial frequency up-shift or decreasing the chirp rate decreases the final amplitude that can be reached during a fixed interaction time.

% section on: experimental examples

\section{Experimental Considerations}\label{section:experimental}

Unfortunately, as has been alluded to previously, the PBWA does not have unlimited time to be excited, as deleterous instabilities will eventually destroy wave coherence.  For the 
parameters of interest, the oscillating two-stream (also referred to as 
modulational) instability limits the lifetime of coherent Langmuir waves 
to the ion time-scale, \ie, for times of the order of a few $1/\omega_{i}$.  Although 
it is possible that the growth of this instability may be mitigated somewhat by the use of a 
chirped laser, in this paper we will use as a conservative figure the 
results of Mora \textit{et. al.} \cite{mora_et_al:88} to set the time 
limit during which we can excite a coherent plasma wave suitable for accelerator applications.

For the relatively cold plasmas and moderately intense lasers we consider, it is 
shown in \cite{mora_et_al:88} that the growth rate of the oscillating 
two-stream instability is approximately equal to $\omega_{i}$, and that 
this instability impedes plasma wave excitation and destroys coherence after about $5$ 
$e$-foldings.  Thus, we see that the drive lasers should have time duration 
$T \lesssim 5/\omega_{i}.$  If one chirps  the drive frequency leading to 
a total shift $\delta\omega$ during the autoresonant excitation, then the normalized chirp rate is limited to 
\be
\alpha \gtrsim 0.2 \, \frac{\omega_p}{\omega_i} \frac{\delta\omega}{\omega_p},
\ee
or approximately 
$\alpha \gtrsim 2.3 \times 10^{-3} \, \delta\omega/\omega_p$ for 
singly-ionized Helium.  Below, we choose two experimentally relevant parameter 
sets, one corresponding to a $10 \mbox{ }\mu\mbox{m} \mbox{ } \mbox{CO}_2$ laser; the other, to a $800 \mbox{ nm}$ Chirped Pulse Amplification (CPA) \cite{strickland_mourou:85, maine_et_al:88} Ti:Sapphire laser system.  We demonstrate how, beginning with the laser frequency above the linear resonance and then slowly decreasing it, one can robustly excite plasma waves to amplitudes larger than the cold, linear wave-breaking limit in times commensurate with onset of the oscillating two-stream instability.

\subsection{$\mbox{CO}_2$ Laser at $10 \mbox{ }\mu\mbox{m}$}\label{subsection:co2}

\begin{figure}
  \centering
  \includegraphics[scale=1]{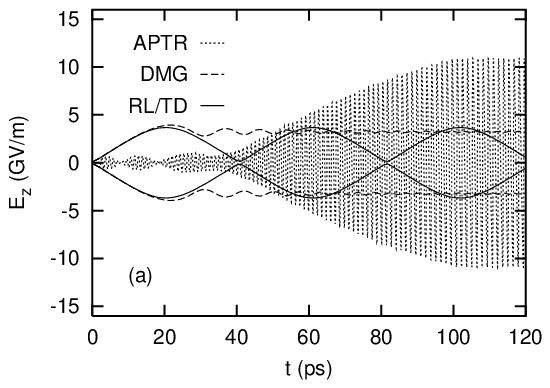}
  \includegraphics[scale=1]{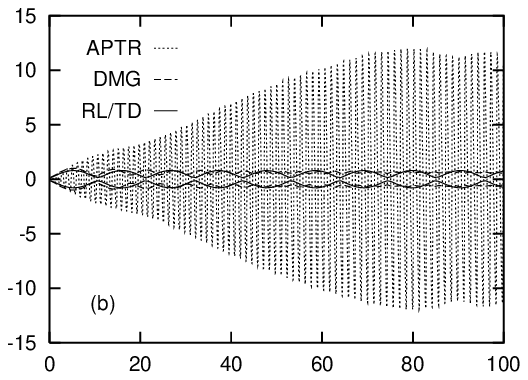}
  \includegraphics[scale=1]{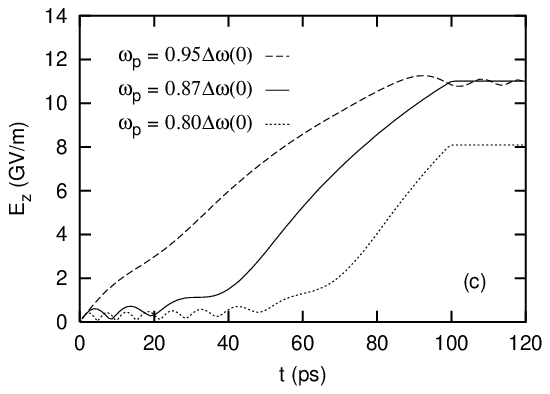}
  \includegraphics[scale=1]{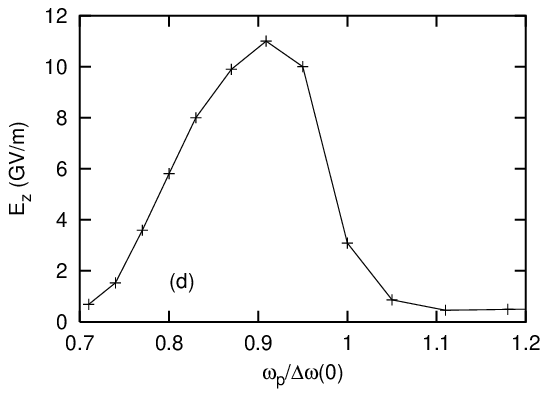}
  \caption{Plasma wave excitation for a 10$\mu$m CO$_2$ laser with 
	intensity $2.7\times 10^{14}\textrm{W/cm}^2$ ($\epsilon = 0.02$).  
	In (a), the APTR case has $\omega(0) = 1.15\omega_p$, and $\omega 
	= \omega_p$ at $t\approx40$ps $(\alpha = 0.00065)$.  Total chirp 
	is 1.5\% of laser frequency.  For comparison, we include the 
	envelopes with no chirp, and DMG chirp starting on resonance.  (b) is the same as (a), with
	$\omega_p$ changed by 10\%.  In (c), we show uniform accelerating fields via APTR for variations 	in $\omega_p$ of order 10\%.  (d) demonstrate robust field excitations of 5-10GeV/m for 
	density errors of order $\pm20\%.$}	\label{fig:UCLA}
\end{figure}
 
We consider parameters roughly corresponding to the most recently published UCLA upgrade \cite{clayton_et_al:98}.  We assume two pulses of duration $T = 100 \mbox{ ps}$ which enter the plasma at $t = 0,$ central wavelengths near $\lambda = 10 \mbox{ }\mu\mbox{m},$ and normalized intensities $a_1 = a_2 = 0.14,$ corresponding to a normalized drive strength $\epsilon = 0.02,$ so the threshold condition (\ref{eqn:alpha_crit}) implies that the normalized chirp rate should satisfy $\alpha(t) = -\tfrac{d}{dt}\Delta\omega(t)/\omega_p < 0.0009$.  We choose a linear chirp, so that in physical units the beat frequency is given by $\Delta\omega(t) = (\mu_0 + \alpha\,\omega_p t),$ where $\mu_0 = 1.15$ and $\alpha = 0.00065,$ with a total frequency sweep from $\Delta\omega(t = 0) = 1.15$ to $\Delta\omega(t = T) = 0.74$.  For these parameters, we collect the results from simulations integrating the quasi-static equation of motion (\ref{eqn:quasi_eom}) in Fig.~\ref{fig:UCLA}.  \ref{fig:UCLA}(a) demonstrates the excitation of a uniform accelerating field $E_z$ of $10 \mbox{ GV/m},$ which is above the linear cold wave-breaking limit of $ E_0 \approx 8.8 \mbox{ GV/m},$ but below the cold relativistic limit $E_{\text{\tiny{WB}}} \approx 61 \mbox{ GV/m}.$  The total chirp is modest, only about $1.5\%$ of the laser carrier frequency $2\pi c/\lambda.$

For comparison, we also plot the simulated envelopes of the longitudinal field for the resonant RL/TD case 
$\Delta\omega(t) = \Delta\omega(0) = 1$, and for the chirped DMG scheme starting on linear resonance, $\Delta\omega(t) = \left(1 - \alpha\,\omega_p t \right),$ but using the same chirp rate as above.  The resonant case demonstrates the characteristic RL limit of $E_{z} \le E_{\text{\tiny{RL}}} = (16\epsilon/3)^{1/3}E_0 \approx 4.2 \mbox{ GV/m},$ whereas the DMG scheme fails to achieve appreciable dynamic phase-locking, and the final plasma wave amplitude is about the same as in the resonant (unchirped) case.  Using approximately these parameters, UCLA experiments have inferred accelerating amplitudes up to 2.8 GV/m \cite{everett_et_al:94} over short regions of plasma.  More recently, plasma density variations corresponding to $E_z \approx 0.2-0.4\mbox{ GV/m}$ have been directly measured with Thomson scattering \cite{filip_et_al:02}.

Perhaps more important than the higher amplitude field in the autoresonant APTR case, 
is the fact that excitation is very robust with respect to mismatches between  
the beat frequency and the plasma frequency.   In practice, these mismatches inevitably result  from limited 
diagnostic accuracy or shot-to-shot jitter in the plasma or  laser parameters.  
Because one sweeps over a reasonably broad frequency range and one only needs to pass 
through the resonance at some indeterminate point during the chirp history, no precise matching is required, and the exact value of the plasma density need not be accurately known.  This robustness is demonstrated in Fig.~\ref{fig:UCLA}(b)-(d).  Plots (b) and (c) show the longitudinal field profile attained when there are variations in the density, and we see that APTR yields uniform, large amplitude fields over a wide range of densities.  In Fig.~\ref{fig:UCLA}(d), we plot the final accelerating gradient achieved via APTR when we vary the value of $\omega_p$ over a range of $\pm 10\%,$ from its ``design'' value, while keeping the laser parameters fixed.  We see large levels of excitation for a wide range in plasma variation, corresponding roughly to density mismatch/errors up to $20\%.$  Thus, not only is autoresonant plasma wave excitation effective in avoiding saturation from detuning, it also mitigates experimental uncertainties in or shot-to-shot variations of plasma density.

\subsection{Ti:Sapphire Laser at 800 nm}\label{subsection:cpa}

\begin{figure}
  \centering
  \includegraphics[scale=1]{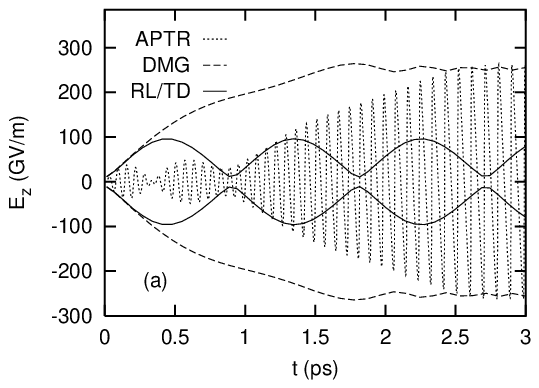}
  \includegraphics[scale=1]{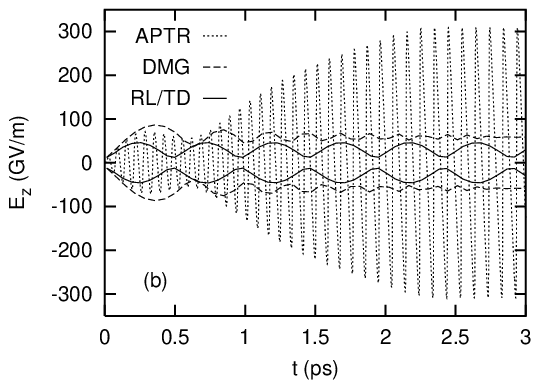}
  \includegraphics[scale=1]{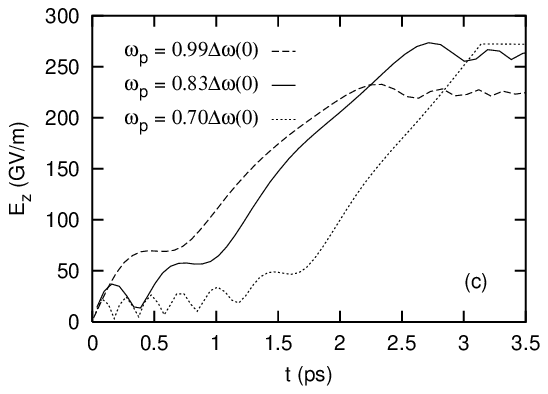}
  \includegraphics[scale=1]{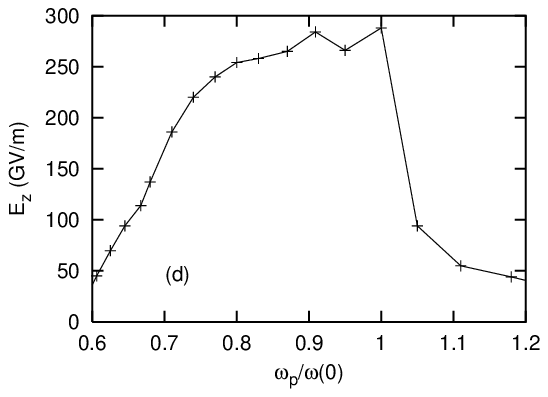}
  \caption{Plasma wave excitation for a $800 \mbox{ nm}$ Ti:sapphire laser with 
	$\bar{\omega}/\omega_p = 25$, intensity $2 \times 10^{17} \mbox{ W/cm}^2$ 
	($\epsilon = 0.09$).  APTR parameters are $\Delta\omega(0) = 1.2$, and 
	$\Delta\omega = 1$ at $t \approx 1.4 \mbox{ ps}$ ($\alpha = 0.00225$).  Total chirp 
	is $3\%$ of the laser frequency.  In (a), we see that the DMG and APTR scheme give
	approximately the same final fields, about three times that for on-resonance.  In (b), we changed
	$\omega_p$ by 10\%, and the excitation for the DMG and the on-resonant scheme drop
	considerably, while APTR maintains its large excitation.  (c) demonstrates that APTR gives
	robust, uniform accelerating fields for wide variations in density, while (d) indicates the insensitivity
	of APTR to errors of $\pm35\%$ in the plasma density, while still yielding large amplitude plasma
	waves.}
						\label{fig:Loasis}
\end{figure}

Here, we analyze a representative case for a Ti:Sapphire CPA laser in a singly-ionized He plasma, with $n_0 = 1.4 \times 10^{18} \mbox{ cm}^{-3},$ so that  $\omega_p/\bar{\omega}= 1/25,$ and a laser duration $T = 3.2 \mbox{ ps},$ chosen to correspond to the modulational instability limit.  If we consider two $1 \mbox{ J}$ pulses compressed to this time duration and focused to a waist of 
$w_0 = 6 \mbox{ }\mu\mbox{m},$ this implies intensities of $I_0 = 2.0 \times 10^{17}\mbox{ W/cm}^2,$ so that, with $\lambda \approx 800 \mbox{ nm},$ we have $a_1 = a_2 = 0.3,$ and $\epsilon = 0.09.$  We choose $\Delta\omega(t = 0) = 1.2$, $\Delta\omega(t = T) = 0.5$, so that $\alpha = 0.0025.$  The resulting plasma wave excitation is shown in Fig.~\ref{fig:Loasis}(a).  Here, we see 
maximum longitudinal electric fields $E_z \approx 260 \mbox{  GV/m},$ corresponding 
to $\sim1.6 \, E_0 \approx 0.25 E_{\text{\tiny{WB}}}.$  For comparison, we also plot the resonant case, for which detuning results in maximum fields corresponding to the familiar RL limit $(16\epsilon/3)^{1/3}E_0 \approx 125 \mbox{ GV/m},$ and the DMG case (starting on resonance), which yields results similar to the APTR case.  The distinction between passing though resonance and starting on resonance can be seen, however, in Fig.~\ref{fig:Loasis}(b), which indicates that a small change in the value of the plasma frequency has a dramatic effect when starting on resonance, but little effect when passing through resonance.  The uniform, robust acceleration fields obtained via APTR are shown in Fig.~\ref{fig:Loasis}(c) for density variations $\pm 20\%$.  Finally, Fig.~\ref{fig:Loasis}(d) shows the robustness of autoresonant 
excitation, for which density imperfections of $\pm 35\%$ have little effect on the accelerating gradients achieved.

% section on: discussion

\section{Discussion: Comparisons, Scalings, and Extensions}\label{section:discussion}

In comparison to other PBWA schemes, including the fixed beat-frequency approach, either at (RL/TD) or below (TSS) linear resonance, the chirped (DMG) scheme, involving downward chirping from resonance, or the non-resonant PBWA, scheme, recently proposed by Filip \textit{et al.} \cite{filip_et_al:02, filip_et_al:03}, involving strongly forced waves at frequency shifts well below resonance in a marginally underdense plasma, the autoresonant/APTR PBWA enjoys a number of advantages, in terms of plasma wave amplitude, robustness, and quality.  In previous sections we have seen how, for given drive laser intensity, autoresonant excitation yields longitudinal fields that can be considerably higher than the RL limit set by relativistic detuning of the plasma wave.  We have also seen how APTR, \ie, slowly sweeping the frequency downward through resonance, provides a much greater degree of robustness to density mismatches, since neither the final amplitude nor frequency of the plasma wave is very sensitive to the precise location of the actual linear resonance, or to the precise chirp history.

\begin{figure}
  \centering
  \includegraphics[scale=1]{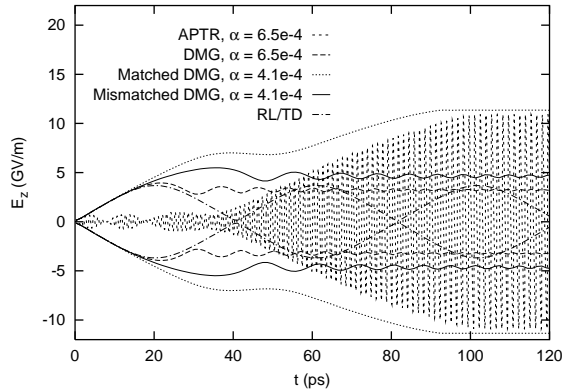}
  \caption{
Comparison of three plasma beat-wave schemes for the $\mbox{CO}_{2}$ laser of Sec.~\ref{subsection:co2}.  Autoresonant/APTR excitation is compared to the DMG scheme with identical chirp, and with a chirp chosen to maximize excitation.  For identical chirp, DMG results in a field of approximately $50 \%$ less amplitude.  The optimized DMG is comparable in amplitude to APTR, but a mismatch in plasma density of $2 \%$ has a sizable effect, unlike the autoresonant/APTR robustness to density fluctuations of order $10 \%$ to $20 \%.$}      \label{fig:accidental_locking}
\end{figure}

Direct comparison with the chirped DMG scheme is slightly more complicated, as revealed in Fig.~\ref{fig:accidental_locking}, because both schemes rely on autoresonance.  In principle, given unlimited excitation time in the APTR case, and for any fixed drive laser intensity, the plasma wave can be autoresonantly excited to any amplitude at or below the nonlinear wave-breaking limit by choosing a sufficiently slow chirp rate.  This is not always true for the DMG case starting at resonance, where performance appears to peak at some intermediate chirp rate, with rapid detuning for significantly faster chirp rates, and excessive ringing for significantly slower chirp.  Without time constraints, the autoresonant APTR approach can always produce higher longitudinal fields for the same drive laser intensity or comparable longitudinal fields with smaller intensity.  But in practice,  the time allowed for excitation is inevitably limited, typically by ion instabilities as previously addressed, or even if these are somehow controlled, then by laser scattering or modulational instabilities, or ultimately by Landau or collisional damping of the Langmuir wave or hydrodynamic expansion of the plasma.

If the DMG and the APTR schemes are compared for realistic parameters using the same drive laser intensity and chirp rate, then autoresonant APTR excitation consistently results in larger electric fields for excitation times on the order of few $\omega_{i}^{-1}.$  But if the DMG scheme is started precisely on resonance, but using a slower chirp rate so as to achieve roughly the same final drive frequency as the APTR approach (which started above resonance), then the DMG approach can experience phase-locking over a longer time and can achieve slightly higher fields in the finite excitation time.  Essentially, by using a sufficiently strong drive, chirp rates that are just slow enough to be adiabatic, and an initial beat frequency precisely at resonance, the DMG trajectory can become autoresonantly phase-locked, but without having to waste time by chirping down from some point well above the resonance as is done in the APTR case.  
The catch is that such phase-locking behavior starting from resonance is quite sensitive to the initial conditions, and DMG will fail to achieve persistent phase-locking if the resonance is missed by just a few percent, so robust phase-locking behavior is unlikely to be experimentally reproducible, unless
one begins with the beat frequency safely above resonance and adiabatically sweep through it, rather than trying to start precisely on resonance.

In order to excite larger electric fields at  a given plasma density, autoresonant/APTR PBWA requires either more intense drive lasers or longer excitation times or both.   Even if technologically feasible, moving to larger $a^2$ introduces its own problems (as discussed below), and in any event tends to undercut one of the primary comparative advantage of the PBWA over either the standard short-pulse or the self-modulated LWFA.  Since excitation times in the PBWA tend to be limited by deleterious ion instabilities (oscillating two-stream instabilities and perhaps ion acoustic instabilities) growing on the $\omega_{i}^{-1}$ time-scale, while for fixed drive laser intensity both the maximum chirp rate and the wave-breaking field scale with the (electron-dominated) plasma frequency $\omega_p,$ from the ratio
\be
\frac{\omega_i}{\omega_p} = \left(\frac{m Z_i}{M_i} \right)^{1/2},
\ee
we observe that, for given plasma density, the time available for plasma wave excitation may be increased simply by (singly) ionizing a gas comprised of a higher-mass atomic species.  With heavier ions, one has more time to excite larger fields with the same laser power, or comparable fields with less laser power before ion motion disrupts coherence.  In addition to ionic instabilities, laser instabilities are also of concern.  Raman backscatter, with growth rate $\Gamma_{\text{\tiny{RBS}}} \sim \smallhalf \bar{a} \sqrt{\bar{\omega}\omega_p}$ and Raman forward scatter, with growth rate $\Gamma_{\text{\tiny{RFS}}} \sim \bar{a}\tfrac{\omega_p}{\bar{\omega_{\phantom{p}}}} \tfrac{1}{\sqrt{8}\gamma_0},$ as well as related 2D self-focusing or modulational instabilities \cite{esarey_et_al:96}, are almost certainly not negligible for parameter regimes of interest, and their effects should be investigated in a more realistic simulations.  There is some hope even that Raman scattering might be turned to our advantage: it may also be possible to exploit autoresonant excitation in counter-propagating or colliding-pulse schemes \cite{shvets_et_al:99, malkin_et_al:00, shvets_fisch:01}, where the enhanced growth of the plasma waves associated with Raman scatter may allow shorter excitation times for fields approaching the wave-breaking limit.

Strictly speaking, our simplified dynamical model has demonstrated the robustness of autoresonant excitation only with respect to global density mismatches, and to small changes in the global chirp rate, drive intensity, or initial detuning.  But we believe that some of this robustness should persist in the presence of  moderate spatio-temporal variations and non-uniformity within the plasma and of complex spatio-temporal structure and dynamics in the lasers, features which are not captured by the idealized model used here.  Effects such as ionization dynamics, thermal fluctuations, hydrodynamic expansion, electron and ion plasma instabilities, ponderomotive blow-out, and of course the nonlinear plasma excitation itself all lead to highly nontrivial plasma density profiles that vary in time and space and change the local conditions for resonance.  Likewise, diffraction, tunneling ionization and the resulting ionization-induced depletion, refraction, and blue-shifting, as well as nonlinear focusing effects due to inhomogeneities and nonlinear back-action of the plasma wave on the EM fields, as well as Raman, self-modulation, and other instabilities, all can lead to appreciable distortion of the driving laser fields during the excitation process.  The importance of these details for autoresonant wake excitation need to be further studied in the context of more detailed laser-plasma models.

Any such variation, inhomogeneity, or complex dynamical structure in the evolving laser fields or plasma is likely to prove deleterious in the absence of autoresonance with adiabatic passage through resonance, drastically reducing wake excitation and uniformity.  Longitudinal fields in some spatial regions may be resonantly excited to moderate or large amplitudes while other regions may be mismatched in density and experience very little excitation, or may be first excited and then subsequently de-excited as a result of phase slippage and energy exchange back to the lasers.  As a result,we expect the final plasma wave to be a highly irregular accelerating structure.  But our autoresonant approach enjoys an intrinsic insensitivity and persistence, due to the local nature of the phase-locking.  Provided only that the magnitude and scales for the non-uniformity are such as to allow an eikonal treatment of the waves, we expect \cite{friedland:92} that at each position, the local plasma response will be autoresonantly excited by the drive, based on the local, slowly-varying values of the plasma frequency, drive amplitude, beat frequency, and chirp rate.  Once phase-locked, the local Langmuir wave will grow monotonically to large amplitude, then saturate with very little ringing.  Not all spatial regions will reach precisely the same final amplitude, but eventually the wave can grow monotonically more or less everywhere until local saturation ensues, so the final variations should be considerably less than in standard approaches.  Although plausible, given what has been demonstrated in previous analyses of autoresonant phenomena, this expectation should also be verified in more realistic simulations.

We have also neglected all thermal effects, but, in realistic situations, tunneling ionization by and inverse bremstrahlung of the drive lasers will typically result in plasma temperatures of  $T_{e} \gtrsim O(10 \mbox{ eV}).$  Damping of the high phase-velocity plasma waves should remain small, but thermal effects can lead to increased particle-trapping and lower thresholds for wave-breaking as well as induce changes in the Langmuir dispersion relation, while also resulting in some background of random short-scale density fluctuations and electrostatic oscillations which may impact the initial stages of autoresonant phase-locking, which assumes a suitably quiescent plasma.  This too may be further invesitgated via numerical simulation.

The suitability of the excited plasma waves for relativistic particle acceleration depends not only on the magnitude and uniformity of the peak electric fields fields but even more crucially on the uniformity of the phase and phase velocity, and on the degree of phase-locking to the external drive.  The Langmuir phase velocity $v_p$ approximately matches the laser group velocity $v_g,$ so first one should consider variability in the latter.  Diffractive effects can substantially lower the longitudinal group velocity of the laser \cite{esarey_et_al:96}, but once accounted for, additional variation due to the finite bandwidth $\delta \omega \sim \omega_p$ are typically quite small in underdense plasmas, $\tfrac{\delta v_g}{v_g}\sim \tfrac{\omega_p^2}{\bar{\omega}^2}\tfrac{\delta\omega}{\bar{\omega}},$ as are variations due to moderate density variations: $\tfrac{\delta v_g}{v_g} \sim \tfrac{\omega_p^2}{\bar{\omega}^2}\tfrac{\delta n}{n_0}.$  The relative change in the dephasing length $L_{\text{d}}$ (for relativistic electrons with initial velocities very near $c$) is correspondingly small: $\frac{\delta L_{\text{d}}}{L_{\text{d}}} \sim \tfrac{\delta v_g}{v_g}.$

With variations in the group velocity $v_g$ expected to be small, phase coherence of the plasma wave will depend on how closely $v_p$ follows the essentially constant $v_g = \bar{v}_g$ of the laser.  Particle-in Cell (PIC) simulations of Filip \text{et al.} \cite{filip_et_al:03} suggest that for the RL scheme, the effective phase velocity of the nonlinear plasma wave (measured in terms of the progression of the field maximum) can vary appreciably, \ie, $10\%$ to $20\%,$  reflecting phase slippage of the Langmuir wave primarily as a result of ponderomotive blowout and relativistic detuning, while the plasma wave produced in their non-resonant PBWA scheme exhibits substantially less phase slippage.   By working within the QSA, in 1D geometry without transverse density variation, we cannot independently assess any such slippage effects for the present scheme, but  we anticipate that it will be similarly small by virtue of the autoresoant phase-locking.  That is, in the resonant RL/TD scheme, plasma inhomogeneities can lead to accelerating buckets that have a changing phase, so that electrons do not experience a 
constant accelerating field.  With autoresonance, however, the laser can 
phase lock to a range of densities, creating an accelerating field of uniform phase that is everywhere directly related to the local phase of the driving beat wave.

An appealing  feature of both the non-resonant PBWA and autoresonant PBWA is this ostensible ability to phase-lock the plasma wave to the beat-wave of the applied drive lasers, with the implied hope that the electron injection, whether based on external cathode injection \cite{clayton_serafini:96} or internal optical injection \cite{esarey_et_al:97, schroeder_et_al:99} can also be phase-locked to the same lasers.  Because of its potential importance, this phase-locking deserves careful investigation.  An obvious worry with the non-resonant PBWA is that entrainment is achieved by brute force, in contrast to our self-trapping using adiabatic passage through resonance.  Because of the intrinsic inefficiency of non-resonant forcing, it must rely on large driving fields and denser plasma to achieve large accelerating fields.  But increasing $\omega_p$ and $a^2$ also increases group-velocity dispersion, which can inhibit phase-locking to an externally-known reference phase, and increases the growth rates for Raman and self-focusing instabilities which can modulate the laser envelope.  One runs the risk of turning the non-resonant PBWA into a self-modulated LWFA.  That is, such modulation can actually enhance the production of plasma waves, but the the plasma response can then become entrained not to the initial laser envelope as applied to the plasma, with a prescribed shape and phase, but to the envelope after it has undergone some uncontrolled nonlinear evolution and modulation.

One must also carefully distinguish phase-locking from frequency-locking, however much all forms of oscillatory entrainment  tend to be conflated under the former name.  Perfect phase-locking implies perfect frequency-locking, and conversely (at least up to some constant but perhaps unknown phase), but  in the case of only partial or imperfect entrainment, it is possible to achieve good frequency-locking without adequate phase-locking, or the converse.  Whether the relative error in the phase-matching or in frequency-matching between driving and driven oscillation is greater depends on whether the Fourier content of fluctuations in the phase is primarily at higher or lower frequencies than the drive frequency itself.

For the purpose of matched particle injection, it is phase-locking which is desired, yet some caution is warranted in claims of true phase-locking in either the non-resonant or autoresonant PBWA.  In any frequency-locked PBWA scheme, the nonlinear frequency of the Langmuir wave may be closely entrained to the precisely known drive frequency, but this does not necessarily imply that the absolute phase of the Langmuir wave may be precisely known.  As a mechanism for phase-locking in the non-resonant PBWA,  the authors  appeal to the claim that an harmonic oscillator, strongly driven off resonance, remains synchronous with the driving force, and then suggest that this should extend to nonlinear oscillators.  But this intuition holds only for damped linear oscillators after the transient is allowed to decay.   If a linear oscillator, with natural frequency $\omega_n,$ and negligible damping on the observational time-scales, is forced by a constant-amplitude, sinusoidal drive with frequency $\omega_d < \omega_n,$ then, independent of the strength of the drive, the driven oscillator will exhibit  persistent, oscillating variations in phase relative to the drive phase $\psi_d = \omega_d t + \psi_d(0)$ which are $\sim O(\omega_d/\omega_n),$ \ie, only first-order in their frequency ratio.  There is no obvious reason to expect in general that nonlinearities will improve matters.

\begin{figure}  
  \centering
   \includegraphics[scale=1]{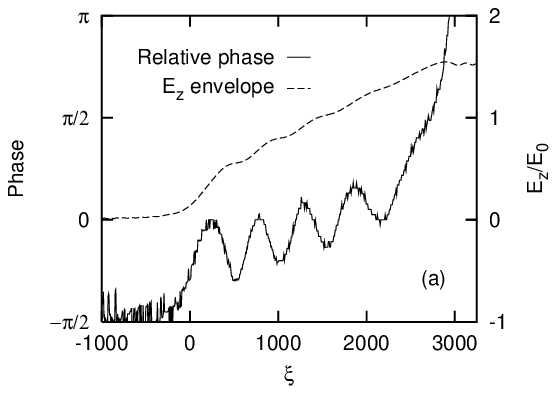}
  \includegraphics[scale=1]{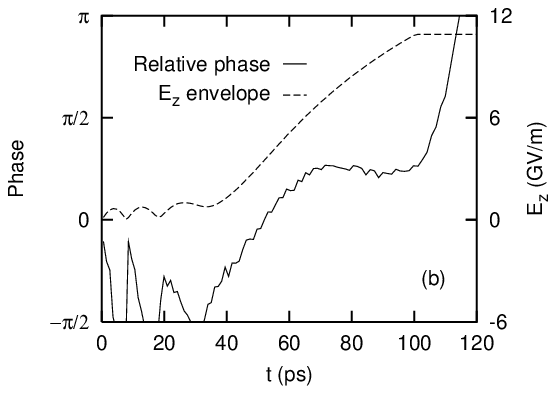}
  \caption{Evolving phase difference between the maxima of the laser beat 
        and the longitudinal electric $E_z$ (which may differ from $\Psi$ 
        in the text by a constant).  Plot (a) has $\epsilon=0.005$, 
        $\alpha = 0.0001$, with the phase slowly evolving during 
        autoresonance for $0 < \xi < 2700,$ where $E_z$ saturates.  (b) 
        uses the CO$_2$ laser parameters of Fig. \ref{fig:UCLA}, 
        which demonstrates nearly constant phase locking during 
        autoresonant excitation for $40<\xi<100$, permitting good control 
        of electron injection.}\label{fig:phase_error}
\end{figure}

In the autoresonant case, we may also encounter seemingly good frequency-locking either with comparably good or disappointingly inferior phase-locking, as shown in Fig.~\ref{fig:phase_error}, where we plot, for two of our previous examples, the phase lag between the beat-wave of the lasers and the plasma wave, numerically estimated by the offset between corresponding relative maxima.  Fig.~\ref{fig:phase_error}(a) exhibits oscillating phase variation which is slowly varying compared to $\omega_p,$ but of order $\lesssim \pi/2$ in absolute (not relative) magnitude, even within the fully nonlinear regime of autoresonance prior to saturation.  On any one shot, an injected electron bunch, if sufficiently relativistic (with electron trajectories $\xi_{e}(\tau) \approx \xi_{e}(0)$ ) and sufficiently short (bunch length $\Delta z_e \ll c\beta_p/c$), would experience uniform acceleration and little additional energy spread.  But the exact phase of the maximal accelerating field could not be reliably predicted from one shot to the next, so there would be considerable shot-to-shot variation in average energy gain of the bunch and no way to reproducibly exploit the peak field, even if the arrival time of the bunch was perfectly correlated with the known beat-phase of the drive lasers in vacuum.

In contrast, in the second case of Fig.~\ref{fig:phase_error}(b), as the oscillator passes through resonance,  the plasma wave phase tightly locks with the drive phase and remains so until adiabaticity is lost and the excitation saturates.  In this case, phase-locked injection scheme could reliably place electron bunches near the maximum acceleration gradient.  It appears that the extent of this phase error is determined  by how deeply the oscillator phase is trapped in its effective potential, which improves with greater drive strength, slower chirp rate, and initial detuning further above resonance.

The most obvious drawback to the autoresonant APTR scheme, in comparison to the standard RL/TD approach, is the added complication of a chirped drive.  But the total chirp required is modest, on the order of only $\omega_p$ in order to guarantee, in the face of some uncertainty as to the density, adiabatic passage through resonance from above.  In typical strongly underdense plasmas this corresponds to only a few percent of the laser carrier frequency.  In solid-state CPA laser systems, this should be relatively easy to achieve, because the laser pulse already undergoes optical stretching/chirping and compression/de-chirping in a series of gratings, and the gain bandwidth of the crystals is intrinsically large ($\sim 20\%$ near $\lambda \sim 1 \mbox{ }\mu\mbox{m}$).  One need only adjust the CPA optics so as to only partially re-compress the final chirped, amplified pulse, in order to leave some residual chirping.

In $\mbox{CO}_2$ lasers (or other gas laser systems), the gain bandwidths tend to be narrower, but can effectively be increased by operating at sufficiently high pressures to doppler-broaden the rotational lines into an overlapping quasi-continuum.  If the final intensities and fluences needed are below the damage threshold for the required optics, one can simply add a pair of gratings similar to those used in CPA systems in order to chirp one of the amplified beams \cite{treacy:69}.  If the final intensity is too large, then with somewhat more difficulty one might arrange a multi-stage system, where an initial seed pulse is chirped in this manner before passing through the final amplifier stage, involving a gas under sufficient pressure to cover the bandwidth of the chirped seed.  More exotically, one might imagine using nonlinear optical effects in a gas or plasma cell to achieve the required frequency shifts.

% conclusions

\section{Conclusion}\label{section:conclusion}

We have introduced a straightforward, seemingly minor, modification of the DMG scheme for the chirped-pulse PBWA, based on the nonlinear phenomenon of autoresonance with adiabatic passage through resonance (APTR), which nevertheless enjoys certain advantages over previous approaches.  Rather than starting at the linear resonance and chirping downward at some intermediate rate expected to match, on average, the beat frequency to the plasma wave frequency corresponding to the growing value of the plasma wave amplitude, we start above resonance and sweep the beat frequency downward past the resonance, at any sufficiently slow chirp rate, such that the plasma wave frequency beat frequency automatically self-locks to the drive frequency, and the plasma wave amplitude automatically adjusts itself consistent with this frequency.  This autoresonant excitation achieves higher plasma wave amplitudes at moderate laser intensities, and, most importantly, appears to be much more robust to inevitable uncertainties and variations in plasma and laser parameters.  Preliminary analysis has been performed within a simplified analytic and numerical model, and wake excitation has been studied using realistic parameters for Ti:Sapphire and $\mbox{CO}_2$ laser systems.  The results are very encouraging, and warrant extending investigation to higher-dimensional geometries, more realistic plasma inhomogeneities, and self-consistent laser evolution, via numerical solution with fluid and particle-in-cell (PIC) codes.

% acknowledgements

\section*{Acknowledgements}\label{section:acknowledgements}

This research was supported by the Division of High Energy Physics, U.S. Department of Energy, by DARPA, U.S. Department of Defense, and by the Israel Science Foundation.

% bibliography 

\bibliographystyle{unsrt}
\bibliography{pbwa_library}

% end of text

\end{document}